\documentclass[%
reprint,
superscriptaddress,
 amsmath, amssymb, aps, longbibliography,prx
]{revtex4-2}

\usepackage{graphicx}
\usepackage{dcolumn}
\usepackage{bm}
\usepackage{comment}

\usepackage{amsmath}
\usepackage{amsfonts}
\usepackage{amssymb}
\usepackage{bm}
\usepackage{graphicx}
\usepackage[T1]{fontenc}
\usepackage{textcomp}
\usepackage{enumitem}
\usepackage{cancel}
\usepackage{bbold}
\usepackage{esvect}
\usepackage{commath}
\usepackage[most]{tcolorbox}
\usepackage{lipsum, babel}
\usepackage{comment}
\usepackage{lineno}

\usepackage{svg}

\begin{document}

\nolinenumbers 
\linenumbersep 3pt\relax

\setlength\columnsep{25pt}

\title{Superballistic conduction in hydrodynamic antidot graphene superlattices}

\author{Jorge Estrada-\'{A}lvarez}
\thanks{Contributed equally to this paper}
\affiliation{
GISC, Departamento de Física de Materiales, Universidad Complutense, 28040 Madrid, Spain.
}%

\author{Juan Salvador-S\'{a}nchez}
\thanks{Contributed equally to this paper}
\affiliation{%
 Nanotechnology Group, USAL--Nanolab, Departamento de Física Fundamental, University of Salamanca,\\
 Plaza de la Merced, Edificio Triling\"{u}e, 37008, Salamanca, Spain.
}%

\author{Ana P\'{e}rez-Rodr\'{i}guez}%
\affiliation{%
 Nanotechnology Group, USAL--Nanolab, Departamento de Física Fundamental, University of Salamanca,\\
 Plaza de la Merced, Edificio Triling\"{u}e, 37008, Salamanca, Spain.
}%

\author{Carlos S\'{a}nchez-S\'{a}nchez}%
\affiliation{%
 Nanotechnology Group, USAL--Nanolab, Departamento de Física Fundamental, University of Salamanca,\\
 Plaza de la Merced, Edificio Triling\"{u}e, 37008, Salamanca, Spain.
}%

\author{Vito Cleric\`{o}}%
\affiliation{%
 Nanotechnology Group, USAL--Nanolab, Departamento de Física Fundamental, University of Salamanca,\\
 Plaza de la Merced, Edificio Triling\"{u}e, 37008, Salamanca, Spain.
}%

\author{Daniel Vaquero}%
\affiliation{%
 Nanotechnology Group, USAL--Nanolab, Departamento de Física Fundamental, University of Salamanca,\\
 Plaza de la Merced, Edificio Triling\"{u}e, 37008, Salamanca, Spain.
}%

\author{Kenji Watanabe}
\affiliation{
Research Center for Functional Materials, National Institute for Materials Science\\
 1-1 Namiki, Tsukuba 305-0044, Japan.
}%

\author{Takashi Taniguchi}
\affiliation{
International Center for Materials Nanoarchitectonics,  National Institute for Materials Science\\
 1-1 Namiki, Tsukuba 305-0044, Japan.
}%

\author{Enrique Diez}%
\affiliation{%
 Nanotechnology Group, USAL--Nanolab, Departamento de Física Fundamental, University of Salamanca,\\
 Plaza de la Merced, Edificio Triling\"{u}e, 37008, Salamanca, Spain.
}%

\author{Francisco Domínguez-Adame}
\affiliation{
GISC, Departamento de Física de Materiales, Universidad Complutense, 28040 Madrid, Spain.
}%

\author{Mario Amado}%
\affiliation{%
 Nanotechnology Group, USAL--Nanolab and IUFFyM, Departamento de Física Fundamental, University of Salamanca,\\
 Plaza de la Merced, Edificio Triling\"{u}e, 37008, Salamanca, Spain.
}%

\author{Elena D\'{i}az}
\email{elenadg@fis.ucm.es}
\affiliation{
GISC, Departamento de Física de Materiales, Universidad Complutense, 28040 Madrid, Spain.
}%

\date{\today} 
             
\begin{abstract}

Viscous electron flow exhibits exotic signatures such as superballistic conduction. In order to observe hydrodynamics effects, a 2D device where the current flow is as inhomogeneous as possible is desirable. To this end, we build three antidot graphene superlattices with different hole diameters. We measure their electrical properties at various temperatures and under the effect of a perpendicular magnetic field. We find an enhanced superballistic effect, suggesting the effectiveness of the geometry at bending the electron flow. In addition, superballistic conduction,  which is related to a transition from a non-collective to a collective regime of transport, behaves non-monotonically with the magnetic field. We also analyze the device resistance as a function of the size of the antidot superlattice to find characteristic scaling laws describing the different transport regimes. We prove that the antidot superlattice is a convenient geometry for realizing hydrodynamic flow and provide valuable explanations for the technologically relevant effects of superballistic conduction and scaling laws. 

\end{abstract}

\maketitle

%
%

\section{Introduction} \label{sec:level1}

Collisions against impurities and phonons dominate electron-electron collisions in ordinary metals in most cases. In 1963, however, Gurzhi claimed that a decrease in the electrical resistance with increasing temperature might appear in ultra-clean metals at moderate temperatures
~\cite{Gurzhi1963}. This author attributed the phenomenon to the realization of a particular transport regime, where highly correlated electrons behave collectively in a similar way to molecules in conventional viscous fluids~\cite{Gurzhi1968}.  
Indeed, the increase in temperature favors electron-electron collisions, such that the boundary scattering is less efficient,  enhancing the electrical current~\cite{effects_of_electron_electron_scattering_in_wide_ballistic_microcontacts, electron_electron_scattering_and_magnetoresistance_of_ballistic_microcontacts, ballistic_flow_of_two_dimensional_interacting_electrons, hall_effect_in_a_ballistic_flow_of_two_dimensional_interacting_particles}. 
This effect, also known as superballistic conduction, constitutes one of the archetypal hydrodynamic signatures~\cite{Polini2020,Narozhny2022,Varnavides2023,Bandurin2016,Bandurin2018,visualizing_poiseuille_flow_of_hydrodynamic_electrons,Kumar17,imaging_hydrodynamic_electrons_flowing_without_landauer_sharvin_resistance}.
Since the collective motion of electrons leads to a resistance below the ballistic limit, superballistic conduction is a convenient property for low-power consumption devices~\cite{how_electron_hydrodynamics_can_eliminate_the_landauer_sharvin_resistance}. A reduction of the resistance by up to $16\%$ has been reported in nanoconstrictions~\cite{Kumar17,boundary_mediated_electron_electron_interactions_in_quantum_point_contacts,higher_than_ballistic_conduction_of_viscous_electron_flows} and up to $4\%$ in crenelated channels~\cite{geometric_control_of_universal_hydodynamic_flow_in_a_two_dimensional_electron_fluid}. 

Another well-recognized hydrodynamic signature is the experimental formation of Poiseuille flow~\cite{visualizing_poiseuille_flow_of_hydrodynamic_electrons}. In conventional fluids, the Poiseuille law is one of the fundamental principles: the resistance scales as $R \propto 1/d_{s}^4$ where $d_s$ is the diameter of a single pipe carrying the fluid~\cite{recherches_experimentales_sur_le_mouvement_des_liquides,an_introduction_to_fluid_dynamics_BATCHELOR,fluid_mechanics_landau_and_lifshitz_course_of_theoretical_physics_volume_6_LANDAU,Gooth2018}. If, instead, the space is filled up with several pipes of diameter $d$, the Poiseuille law reads $R \propto 1/d^2$~\cite{an_introduction_to_fluid_dynamics_BATCHELOR}. 
This situation is the equivalent of an antidot superlattice in a two dimensional~(2D) system. 

In the last decade, the development of 2D materials has multiplied the experimental realization of electron hydrodynamics. In this transport regime, electrons travel long distances, larger than the size of typical devices, before scattering against point defects or phonons. This is the case of ultra-pure PdCoO$_2$~\cite{Varnavides2023,nonlocal_electrodynamics_in_ultrapure_pdcoo2}, Weyl semimetals~\cite{direct_observation_of_vortices_in_an_electron_fluid}, (Al,Ga)As heterostructures~\cite{geometric_control_of_universal_hydodynamic_flow_in_a_two_dimensional_electron_fluid} or graphene~\cite{visualizing_poiseuille_flow_of_hydrodynamic_electrons,Kumar17,Bandurin2016,Bandurin2018}, with improved properties after hexagonal boron nitride (hBN) encapsulation~\cite{boron_nitride_substrates_for_high_quality_graphene_electronics}. However, direct visualization of the hydrodynamic flow often requires a complex microscopy setup~\cite{visualizing_poiseuille_flow_of_hydrodynamic_electrons,imaging_hydrodynamic_electrons_flowing_without_landauer_sharvin_resistance,observation_of_current_whirlpools_in_graphene_at_room_temperature,direct_observation_of_vortices_in_an_electron_fluid}, making it difficult to define a ubiquitous criterion to establish the occurrence of hydrodynamic transport~\cite{visualizing_poiseuille_flow_of_hydrodynamic_electrons,Bandurin2018, transverse_magnetosonic_waves_and_viscoelastic_resonance_in_a_two_dimensional_highly_viscous_electron_fluid, meta_hydrodynamic_routes_to_viscous_electron_flow}. Due to the potential applications~\cite{elimating_the_channel_resistance_in_two_dimensional_systems_using_viscous_charge_flow,how_electron_hydrodynamics_can_eliminate_the_landauer_sharvin_resistance,hydrodynamical_study_of_terahertz_emission_in_magnetized_graphene_field_effect_transistors,terahertz_radiation_from_the_dyakonov_shur_instability_of_hydrodynamic_electrons_in_a_corbino_geometry,Huang2023} in electronic design and the need to easily explore new materials~\cite{Narozhny2022}, it is desirable to look for novel platforms for viscous electron flow. 

In this work, we propose a graphene-based structure with superballistic and hydro\-dynamic effects in the electron transport. We demonstrate that electron trajectories become naturally bent after geometrically engineering the device, which boosts the hydrodynamic signatures. Namely, we build different antidot graphene superlattices to study superballistic conduction as an indicator of collective electron flow. We also demonstrate a non-monotonic electrical response as a function of the magnetic field and study scaling laws that resemble the traditional fluid Poiseuille law. Our measurements are perfectly supported by detailed theoretical simulations that offer a better insight into the fundamentals of superballistic conduction. 

%
%

\section{Geometrically engineered devices for hydrodynamics} \label{sec:samples}

\begin{figure}[t!]
\includegraphics[width=0.9\columnwidth]{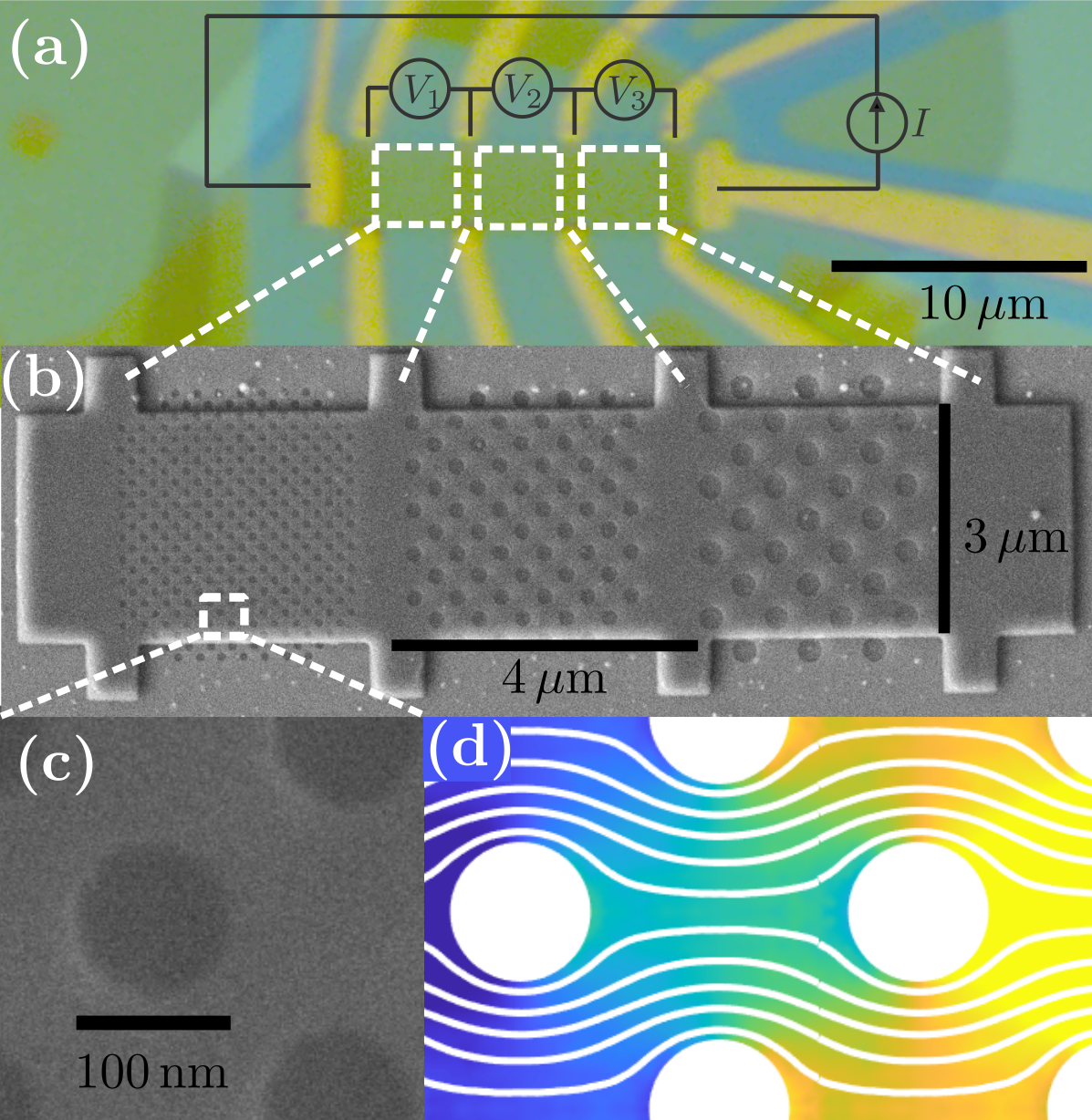}
\caption{Graphene antidot superlattice. (a)~Optical image and schematics of the device. 
(b)~SEM micrograph of an hBN flake showing the same antidot geometry lithographed onto the final device. Three regions can be found, where the antidot diameters are $d = 100, \, 200 $, and $300 \, \rm nm$. The center-to-center antidot distance is $2d$ and they were arranged into a square lattice. (c)~Enlarged SEM micrograph for $d=100\,$nm antidots displaying smooth edges.  (d)~Simulation of the Boltzmann equation, where colors account for the electric potential and streamlines are average electron trajectories. 
}
\label{fig:sample}
\end{figure}

In order to study viscous effects on electron flow, we need to design a 2D device with a current flow that is as inhomogeneous as possible~\cite{Bandurin2016}. Here, we design an optimized anti\-dot superlattice (see Appendix \ref{app:GeometryOptime}) to bend the electron flow and favor hydrodynamics in a fully encapsulated graphene heterostructure.  The latter was fabricated by means of the standard mechanical exfoliation on pristine crystals of hBN and graphite (see Appendix~\ref{app:SamplePrep}). The heterostructure was finally shaped using e-beam lithography into a typical 10--terminal Hall bar with a longitudinal contact-to-contact distance $L=4\,\mu$m, shown in Fig.~\ref{fig:sample}(a). A last step, electron beam lithography and cryo etching~\cite{Clerico2019,EBLvito2020} process were performed in order to define the antidot patterns with smooth edges that ensure almost specular reflection~\cite{Clerico2019,EBLvito2020} [see Fig.~\ref{fig:sample}(b) and (c)]. Three separated regions are shown, consisting of three antidot superlattices of different diameter, namely $d = 100, \, 200 $, and $300 \, \rm nm$. The antidots appear in a square lattice with a center-to-center distance $2d$. Figure~\ref{fig:sample}(d) displays the bent electron trajectories in this geometry.  

Viscous electron flow results from a collective motion of electrons that is expected to reduce the device resistance~\cite{Gurzhi1963,Gurzhi1968}. In particular, superballistic conduction involves a transition from a non-collective to a collective regime of transport. Consequently, we rely on the semiclassical Boltzmann transport equation~\cite{meta_hydrodynamic_routes_to_viscous_electron_flow,ballistic_and_hydrodynamic_magnetotransport_in_narrow_channels} (see Appendix~\ref{app:theoreticalModels}). The latter describes the distribution $g(\bm r, \theta )$ of electrons at position $\bm r$ moving in the direction of $\theta$ as
\begin{equation}
\left( \begin{matrix} \cos \theta \\ \sin \theta \end{matrix} \right) 
\cdot \nabla \left( g- \frac{e V(\bm r)}{\hbar k_F} \right)+ \frac{\partial_\theta g}{l_B} + \frac{g}{l_e}   
+ \frac{g-g_{ee}}{l_{ee}}  = 0 \ ,
\label{BTE}%
\end{equation}
where $k_F = \sqrt{\pi |n|}$ is the Fermi wavenumber for a 2D carrier density $n$ and $g_{ee}$ is given by Eq.~\eqref{eq:gee}. Electrons move under the effect of an electric  potential $V(\bm r )$ and a perpendicular magnetic field ${\bm B}$ with a cyclotron radius $l_B = \hbar k_F / e B$. The Boltzmann equation is solved numerically using a finite element method and its solution is used to compute the drift velocity and the electrical resistance (see Appendix~\ref{app:numericalMethods}).  Together with the collisions against the edges, which we generally consider as specular reflections against a smooth edge~\cite{boundary_conditions_of_viscous_electron_flow}, Eq.~\eqref{BTE} accounts for different mechanisms of electron scattering. First, electron collisions with defects or phonons alter the total momentum of the electrons. We quantify them with the mean free path such that $l_{e}^{-1}=l_{e,\rm{imp}}^{-1}+l_{e,\rm{ph}}^{-1}$. We estimate $l_{e,\rm{imp}} = 700 \, \rm nm$ at low temperatures and $n = 0.3 \times 10^{12} \, \rm cm^{-2}$, consistent with other nanostructured devices~\cite{ballistic_transport_in_graphene_antidot_lattices_2,ballistic_transport_in_graphene_antidot_lattices}. We also take into account phonon scattering at higher temperatures with mean free path $l_{e,\rm{ph}}$~\cite{phonon_mediated_room_temperature_quantum_hall_transport_in_graphene,acoustic_phonon_scattering_limited_carrier_mobility_in_two_dimensional_extrinsic_graphene} (see Appendix~\ref{app:mobility}). Second, electron-electron collisions conserve the total momentum, and the corresponding mean free path $l_{ee}$ can be computed for graphene~\cite{Kumar17,quantum_theory_of_the_electron_liquid}. In this work, we consider the tomographic description~\cite{viscosity_of_two_dimensional_electrons,tomographic_dynamics_and_scale_dependent_viscosity_in_2D_electron_systems,linear_in_temperature_conductance_in_electron_hydrodynamics,collective_modes_in_interacting_two_dimensional_tomographic_fermi_liquids,anomalously_long_lifetimes_in_two_dimensional_fermi_liquids,nonequilibrium_relexation_and_odd_even_effect_in_finite_temperature_electron_gases} beyond Callaway's ansatz~\cite{Callaway1959}, and we define two relaxation rates for the even and odd-parity modes in the angular expansion of $g$, $l_{ee}^\mathrm{even} = l_{ee}$ and $l_{ee}^\mathrm{odd} \gg l_{ee}$. Electron-electron collisions together with the antidot geometry lead to superballistic conduction, beyond the description of Matthiessen's rule.

%
%

\section{Results and discussion} \label{sec:results}

\subsection{Enhanced superballistic effect}
\label{superballistic}

\begin{figure*}[ht!]
\includegraphics[width=18cm]{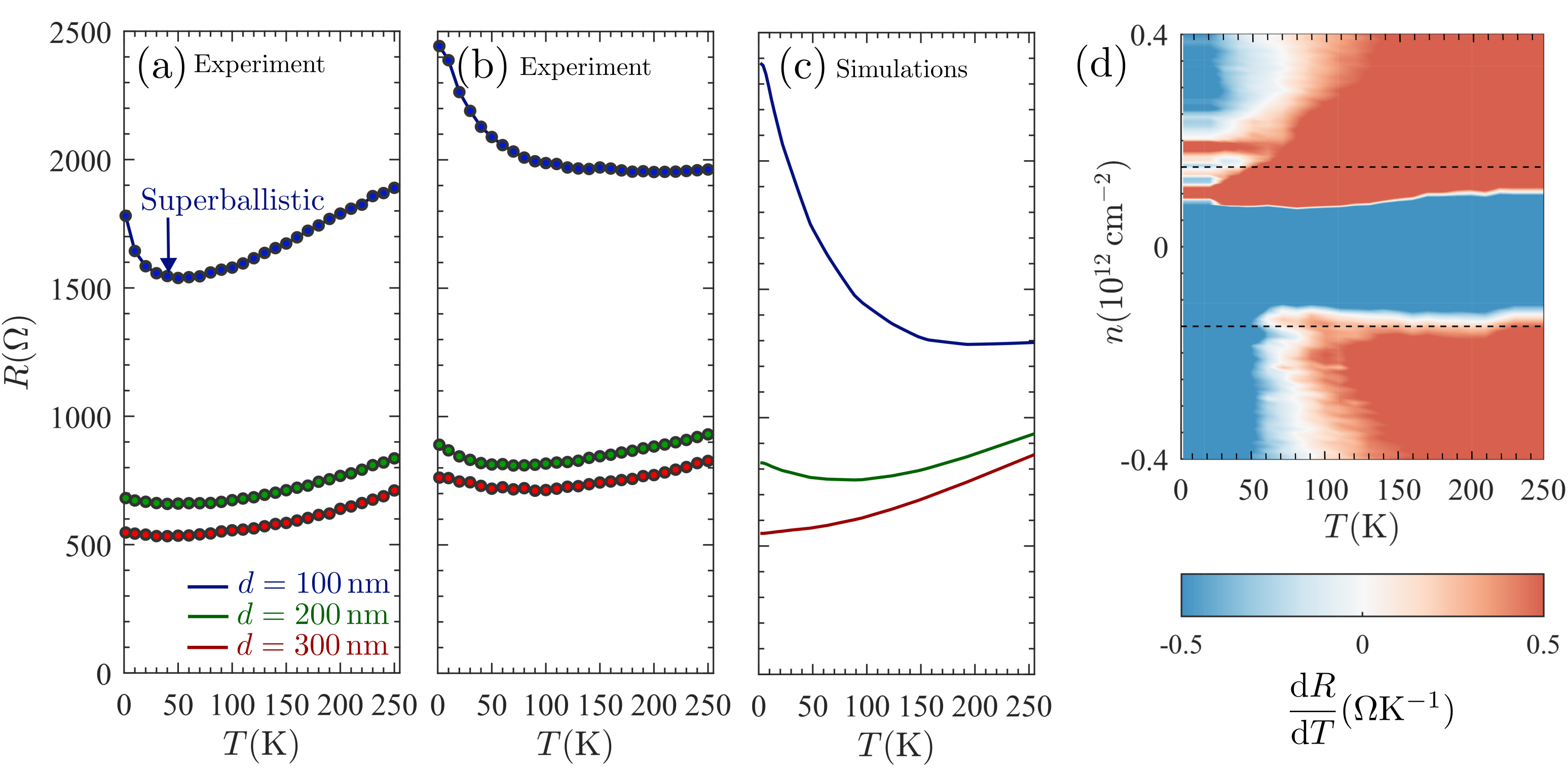}
\caption{Enhanced superballistic conduction. Experimental resistance in the three regions of the antidot lattice with $d = 100, \, 200 $, and $300 \, \rm nm$ as a function of temperature for (a)~electrons at $n = 0.3 \times 10^{12} \, \rm cm^{-2}$ and (b)~holes at the same density. (c)~Theoretical resistance calculated by Boltzmann equation simulations. (d)~Experimental magnitude of $\text{d}R/\text{d}T$ with increasing temperature and various densities of carriers for the $d = 200 \, \rm nm$ region.}
\label{fig:superballistic}
\end{figure*}

Experimental results shown in Figs.~\ref{fig:superballistic}(a) and~(b) clearly exhibit a decrease in the electrical resistance below its ballistic limit at low temperatures in the three regions with antidots.  Thus, our results demonstrate the superballistic effect and support the effectiveness of the antidot superlattice at bending the electron flow.
For a given temperature, there are more electron-electron collisions for lower carrier densities, favoring the collectivization of the electron flow and the superballistic effect. Therefore, consistently with previous works~\cite{Kumar17}, we focus hereafter on the study of lower carrier densities $n = 0.3\times 10^{12} \, \rm cm^{-2}$.
Figure~\ref{fig:superballistic}(c) shows the predictions of the Boltzmann equation for the same antidot geometries. In particular, we discuss in Appendix~\ref{app:chargeInhomogenity} the influence of possible charge inhomogeneity effects to explain the better agreement with the experiment for hole conduction. Also, the latter shows enhanced superballistic conduction in the region of $d=100 \, \rm nm$ that may survive near room temperature. This result, different from the one observed in the regions of $d=200 \, \rm nm$ and $300 \, \rm nm$, had not been reported in nanoconstrictions~\cite{Kumar17}. Nevertheless, it is accurately predicted by the Boltzmann equation. Indeed, both the bending of the electron flow and the smaller ratio $d/l_e$ restrain the detrimental impact of phonon scattering. Similar to the formation of current whirlpools recently observed in graphene~\cite{observation_of_current_whirlpools_in_graphene_at_room_temperature}, our work is consistent with collective electron transport at room temperature as well. This brings in a paradigm where the technological advantages of reduced electrical resistance can be further exploited. 

For completeness, Fig.~\ref{fig:superballistic}(d) monitors the super\-ballistic effect by means of $dR/dT$ in the region of $d=200 \, \rm nm$ when the density of carriers $n$ and $T$ are varied. 
The region close to the charge neutrality point, limited by dashed lines in Fig.~\ref{fig:superballistic}(d), also exhibits negative $\text{d}R/\text{d}T$ due to thermal excitations that lead to a Dirac plasma regime~\cite{giant_magnetoresistance_of_dirac_plasma_in_high_mobility_graphene}. The latter shall not be confused with superballistic conduction. 
The reduction of resistance for $d = 100\, \rm nm$ with $n = 0.2 \times 10^{12} \, \rm cm^{-2}$
is estimated to be larger than $20\%$, resulting in a remarkable improvement to those values previously reported in other systems~\cite{geometric_control_of_universal_hydodynamic_flow_in_a_two_dimensional_electron_fluid}. Once again, the latter mostly results as a consequence of the larger bending of the electron trajectories by the antidot superlattice. Contrary to what was expected, the condition  $l_{ee} \ll d \ll l_{e}$ is not essential for the Gurzhi effect to occur~\cite{Gurzhi1963,Gurzhi1968}. Notice that $l_{ee} $ is indeed very large in the low-temperature limit. For example, in our experiment, we demonstrate superballistic conduction even at $T = 50 \, \rm K$ when $l_{ee} \approx 1100 \, \rm nm$  while $l_{e} \approx 600 \, \rm nm$. However, in the ballistic regime, electron-electron collisions still lead to a positive contribution to the current due to electrons scattered away from the edges. This collective regime can be considered as a precursor of hydrodynamic flow~\cite{effects_of_electron_electron_scattering_in_wide_ballistic_microcontacts, electron_electron_scattering_and_magnetoresistance_of_ballistic_microcontacts, ballistic_flow_of_two_dimensional_interacting_electrons, hall_effect_in_a_ballistic_flow_of_two_dimensional_interacting_particles}. Furthermore, such a result is supported by our simulations that include the particular geometrical details beyond the approximated anti-Matthiessen rule~\cite{Kumar17,higher_than_ballistic_conduction_of_viscous_electron_flows}. This experimental evidence violates the standard expected criteria for collective electron flow, $l_{ee} \ll d$ and $l_{ee} \ll l_e$~\cite{Polini2020, Narozhny2022}, so another reasoning must be developed. We conclude that not only the values of $l_{e}$ and $l_{ee}$ determine the collectivization of the electron flow but mostly their decrease rate with temperature since $\text{d}R/\text{d}T=(\partial R/\partial l_e)(\partial l_e/\partial T) + (\partial R/\partial l_{ee}) (\partial l_{ee}/\partial T)$. Thus, the existence and the magnitude of the superballistic effect mainly depend on such rates. 

In our theoretical study, we also paid attention to the relevance of the boundary conditions to reproduce the experimental results with our model~\cite{boundary_conditions_of_viscous_electron_flow} (see Appendix~\ref{app:boundaryScattering}). Similarly to previous works~\cite{geometric_control_of_universal_hydodynamic_flow_in_a_two_dimensional_electron_fluid}, we can conclude that the demonstrated superballistic effect is almost universal since it relies on the geometry of the device and not on the particular considered edge scattering mechanism. We also check the superballistic conduction in another device where the experimental measurements agree with these results (see Appendix \ref{app:reproducibility}). Most importantly, in the additional device, we demonstrate that there is no decrease in the resistance in an area where no antidots were fabricated. This further confirms that geometrical engineering is responsible for the observed superballistic conduction. 

\begin{figure*}[ht!]
\includegraphics[width=18cm]{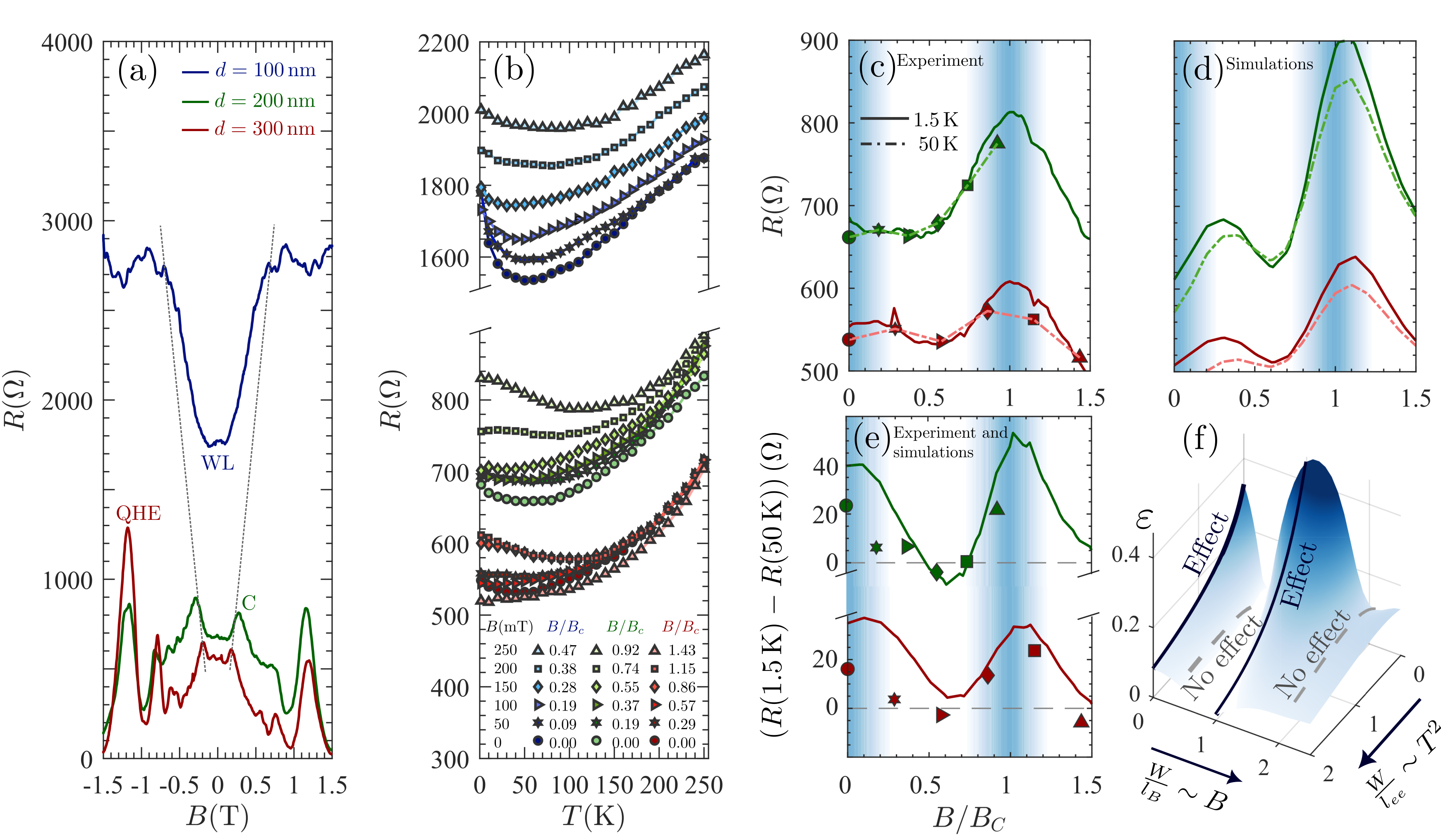}
\caption{Intermittent superballistic conduction. (a)~Longitudinal resistance for the three regions at $n = 0.3 \times 10^{12} \, \rm cm^{-2}$ and $T = 1.5 \, \rm K$ shows weak localization, commensurability effect, and oscillations associated with the quantum Hall effect. (b)~Superballistic effect for several magnetic fields. 
(c)~Experimental measurements and (d)~Boltzmann equation simulations of the resistance as a function of the magnetic field, normalized to the commensurability fields $B_C$ of the peak. (e)~Comparison of results shown in~(c) and~(d), where symbols are experimental observations and lines are simulated values. (f)~Qualitative explanation of the intermittent effect, where $W$ is the typical width of a uniform device.}
\label{fig:intermittent}
\end{figure*}

\subsection{Magnetotransport: Intermittent superballistic effect}

Here we explore the superballistic effect in more detail by considering the electrical response of the antidot lattice in the presence of a perpendicular magnetic field~\cite{measuring_hall_viscosity_of_graphene_electron_fluid, driving_viscous_hydrodynamics_in_bulk_electron_flow_in_graphene_using_micromagnets, negative_magnetoresistance_in_viscous_flow_of_two_dimensional_electrons}. 
First, let us analyze the low-temperature resistance as a function of the applied magnetic field, as shown in Fig.~\ref{fig:intermittent}(a). Evidence of two quantum effects is shown in our measurements: i) the weak localization peak at $B \lesssim 20 \, \rm mT$, due to quantum interference~\cite{quantum_interference_corrections_to_magnetoconductivity_in_graphene} and enhanced by intervalley scattering against the superlattice~\cite{weak_localization_and_transport_gap_in_graphene_antidot_lattices} and ii) the quantum Hall effect, whose peaks flatten for smaller values of $d$, suggesting a prominent role of the antidot geometry. More relevant for the matter of interest are the peaks at the particular field $B_C \approx 1.05\, \hbar \sqrt{\pi n} / e d$ related to the commensurability effect, occurring when the cyclotron radius is commensurate to the antidot lattice size~\cite{visualizing_poiseuille_flow_of_hydrodynamic_electrons,Imaging_viscous_flow_of_the_Dirac_fluid_in_graphene,viscous_transport_and_hall_viscosity_in_a_two_dimensional_electron_system,electron_trajectories_and_magnetotransport_in_nanopatterned_graphene_under_commensurability_conditions,ballistic_transport_in_graphene_antidot_lattices,ballistic_transport_in_graphene_antidot_lattices_2,ballistic_hydrodynamic_phase_transition_in_flow_of_two_dimensional_electrons}. Consequently, these peaks are shifted to higher magnetic fields for decreasing $d$, as indicated by the dotted line in Fig.~\ref{fig:intermittent}(a). For $d=200$ and $300 \, \rm nm$, a region with negative magnetoresistance following this peak is also visible. Negative magnetoresistance was also studied in GaAs with macroscopic defects~\cite{Relevance_of_weak_and_strong_classical_scattering_for_the_giant_negative_magnetoresistance_in_two_dimensional_electron_gases,hydrodynamic_magnetotransport_in_two_dimensional_electron_systems_with_macroscopic_obstacles} with analogous results to those of a Poiseuille flow~\cite{negative_magnetoresistance_in_viscous_flow_of_two_dimensional_electrons}. In the presence of a magnetic field electrons move in circular orbits. Hence, some electrons can propagate without reaching the edges of the device if the cyclotron radius is short enough~\cite{ballistic_hydrodynamic_phase_transition_in_flow_of_two_dimensional_electrons}. This ensures the transition to collective flow. However in the ballistic regime electron have straight trajectories and boundary collisions are more frequent.

Regarding the superballistic conduction, Fig.~\ref{fig:intermittent}(b) shows that the effect arises for some magnetic fields and disappears for others, in an intermittent pattern. Figs.~\ref{fig:intermittent}(c) and (d) show the agreement between the experimental measurements and simulations of the Boltzmann equation and reveal the interplay between the commensurability field and the intermittent pattern as a function of $B/B_C$. For comparison Fig.~\ref{fig:intermittent}(e) presents a combination of data shown in panels~(c) and~(d). Here the strongest resistance reduction reveals not only at zero magnetic field but also at the commensurability condition. Notice the conventional formalism based on the anti-Matthiessen rule~\cite{higher_than_ballistic_conduction_of_viscous_electron_flows,Kumar17} cannot reproduce this intermittent effect, not even qualitatively, so the Boltzmann equation must be used instead. For completeness, the relevant role of the tomographic description in describing electron transport is explored in Appendix~\ref{app:tomographic}. 

In both physical scenarios, the increase in temperature leads to a transition from a non-collective to a more collective electron flow, which favors the superballistic effect [notice the blue solid lines in Fig.~\ref{fig:intermittent}(f)]. In Fig.~\ref{fig:intermittent}(f) we represent $\varepsilon$ as a measurement of the collective nature of the transport in a uniform channel. The latter is estimated as the deviation of the Boltzmann equation simulations~\cite{meta_hydrodynamic_routes_to_viscous_electron_flow} with respect to a fully collective hydrodynamic model ($\varepsilon\approx 0$)~\cite{Bandurin2016,Polini2020}. Although ballistic effects are dominant at low temperatures~\cite{measuring_hall_viscosity_of_graphene_electron_fluid}, a magnetic field different from $B_C$ enhances electron-electron collisions, such that the transport has relevant collective features even at low temperatures. For such fields, the temperature is not the main agent to boost the transition from a non-collective flow to a more collective one and the superballistic effect does not arise [see dashed grey lines in Fig.~\ref{fig:intermittent}(f)]. In conclusion, the intermittent superballistic effect we observe as a function of the applied magnetic field is consistent with previous predictions of collective electron transport~\cite{ballistic_hydrodynamic_phase_transition_in_flow_of_two_dimensional_electrons,meta_hydrodynamic_routes_to_viscous_electron_flow}.

\subsection{Quasi-Poiseuille law}

\begin{figure}[ht!]
\includegraphics[width=\columnwidth]{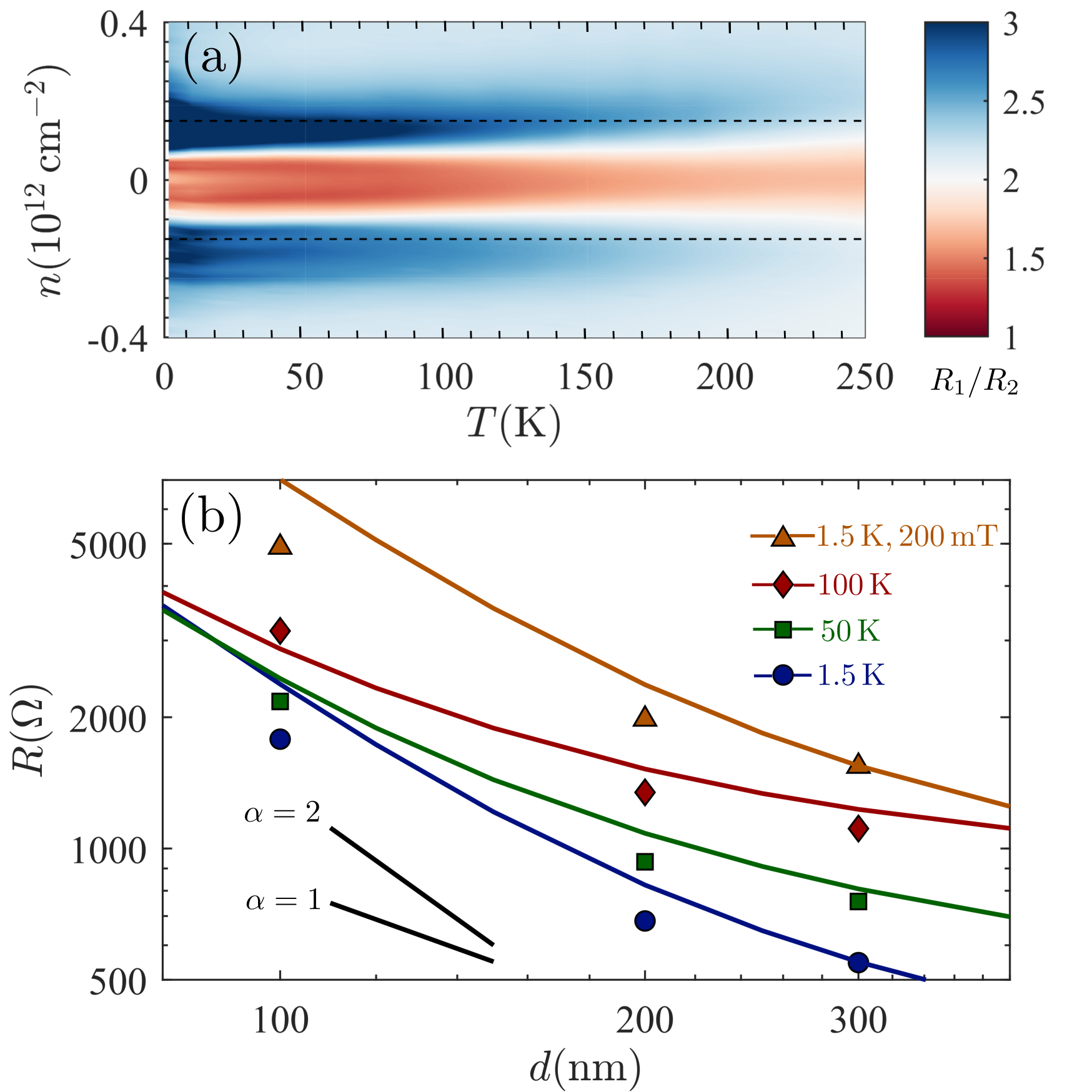}
\caption{Scaling laws. (a) The experimental ratio  between the resistance in the $d = 100$ and $200\, \rm nm$ regions.
(b) Log-log plot for the resistance $R$ versus the antidot size $d$ at $n = 0.3 \times 10^{12} \, \rm cm^{-2}$. Symbols are experimental results and lines are simulations, see the guidelines for the slopes $\alpha = 1$ and $2$.
Curves have been shifted for proper visualization.}
\label{fig:scaling}
\end{figure}
We find a strong dependence of the electrical properties with the antidot size $d$. Let us quantify it by assuming $R \propto 1/d^\alpha$ where $\alpha = - { \partial \log R }/{ \partial \log d}$. For instance, $\alpha = 0$ for diffusive transport, $\alpha = 1$ for the case of the Landauer's formula~\cite{di_ventra_electrical_transport_in_nanoscale_systems,Kumar17}, and
\begin{equation}
\alpha \simeq \frac{2}{1+ \mathrm{v}_F d^2/\beta\nu l_e}\ ,
\label{alphaMainArticle}
\end{equation}
when a hydrodynamic description applies, being $\nu$ the viscosity and $\beta \approx 4.4$ for the antidot geometry (see Appendix~\ref{app:scalingDetail}). Because of the unavoidable inelastic collisions, reducing the viscosity to enter the hydrodynamic regime does not increase $\alpha$, but rather diminishes it. Poiseuille's law for conventional fluids reads $\alpha = 2$~\cite{an_introduction_to_fluid_dynamics_BATCHELOR}, but impurities and phonons affect electron transport and allow for $0 < \alpha < 2$. Therefore, $1 < \alpha < 2$ is assumed as a landmark for electrons~\cite{Gooth2018}, with the largest $\alpha$ arising near the ballistic regime. Indeed, this explains the observations in Fig.~\ref{fig:scaling}(a), where we plot the ratio $R_1 / R_2 \sim 2^\alpha$, being $R_1$ ($R_2$) the resistance in the $d = 100$ ($200 \, \rm nm$) region. The maximum ratio occurs at low temperatures and intermediate densities of carriers, where $l_{e,\rm{imp}}$ is larger in graphene~\cite{Bandurin2016}. We also study the scaling laws in Fig.~\ref{fig:scaling}(b) and compare them with simulations of the Boltzmann equation. The ballistic regime at low temperatures and shorter $d$ gives the largest values of $\alpha$. For increasing temperatures or under the effect of a magnetic field, a transition to a more hydrodynamic regime is favored so that $\alpha$ is reduced. Eventually, a diffusive regime with $\alpha \to 0$ is attained. In summary, the adapted quasi-Poiseuille law gives relevant information about the regime of transport.

%
%

\section{Conclusions} \label{sec:conclusions}

We found that the antidot graphene superlattices are a convenient geometry for the realization of electron hydrodynamics. Indeed, nanostructuring a graphene flake with antidots turns a  material into a \textit{meta-material} showing hydrodynamic response. Certainly, the achievement is supported by the observation of the superballistic conduction with a reduction of the resistance (larger than $20\%$). The antidot geometry bends the electron trajectories so much that the role of edge scattering is overshadowed, leading to an almost universal electron flow~\cite{geometric_control_of_universal_hydodynamic_flow_in_a_two_dimensional_electron_fluid}. 

The observed effects contribute to a better understanding of the so-called hydrodynamic signatures. The solution of the Boltzmann transport equation accounts for the ballistic-hydrodynamic transition, and enables us to study the tomographic dynamics of electrons~\cite{tomographic_dynamics_and_scale_dependent_viscosity_in_2D_electron_systems}. Thereafter, we improve the conventional hydrodynamic description and, in particular, conclude that the anti-Matthiessen rule cannot account for our experimental results~\cite{higher_than_ballistic_conduction_of_viscous_electron_flows}. Particularly, our formalism  based on the Boltzmann equation is crucial to reproduce the intermittent superballistic effect experimentally found as a function of the applied magnetic field. Our description also sheds light on the classification of the hydrodynamic and ballistic transport regimes~\cite{ballistic_hydrodynamic_phase_transition_in_flow_of_two_dimensional_electrons}, by showing that collective regions are achieved in the presence of a magnetic field, in agreement with previous works~\cite{meta_hydrodynamic_routes_to_viscous_electron_flow}. Scaling laws also provide insight into transport regimes. \\

Both the advantages of the antidot superlattices and the feasibility of their fabrication with nanolithography open an avenue to further optimize the geometries, look for novel signatures of viscous electron flow~\cite{negative_differential_resistance_of_viscous_electron_flow_in_graphene}, and explore materials to reduce the resistance in electron devices of technological interest. 

\begin{acknowledgments}

We wish to acknowledge R. Brito, R. Soto, and A. Hamilton for discussions. This work was supported by the “(MAD2D-CM)-UCM” project funded by Comunidad de Madrid, by the Recovery, Transformation and Resilience Plan, and by NextGenerationEU from the European Union, Agencia Estatal de Investigaci\'{o}n of Spain (Grant PID2022-136285NB-C31/C32) and FEDER/Junta de Castilla y León Research (Grant SA106P23). J.~E.-A. acknowledges support from the Spanish Ministerio de Ciencia, Innovaci\'{o}n y Universidades (Grant FPU22/01039). J. S.-S. acknowledges financial support from the Consejería de Educación, Junta de Castilla y León, and ERDF/FEDER. A.~P.-R. acknowledges the financial support received from the Marie Skłodowska Curie-COFUND program under the Horizon 2020 research and innovation initiative of the European Union, within the framework of the USAL4Excellence program (Grant 101034371).

\end{acknowledgments}

\appendix

%
%

%
\setcounter{equation}{0}
\setcounter{figure}{0}
\setcounter{table}{0}
\makeatletter
\renewcommand{\theequation}{S\arabic{equation}}
\renewcommand{\thefigure}{S\arabic{figure}}
\renewcommand{\bibnumfmt}[1]{[#1]} 
\renewcommand{\thetable}{S\arabic{table}}
%

\section{Geometry optimization}
\label{app:GeometryOptime}

\begin{figure}[ht!]
\includegraphics[width=0.8\columnwidth]{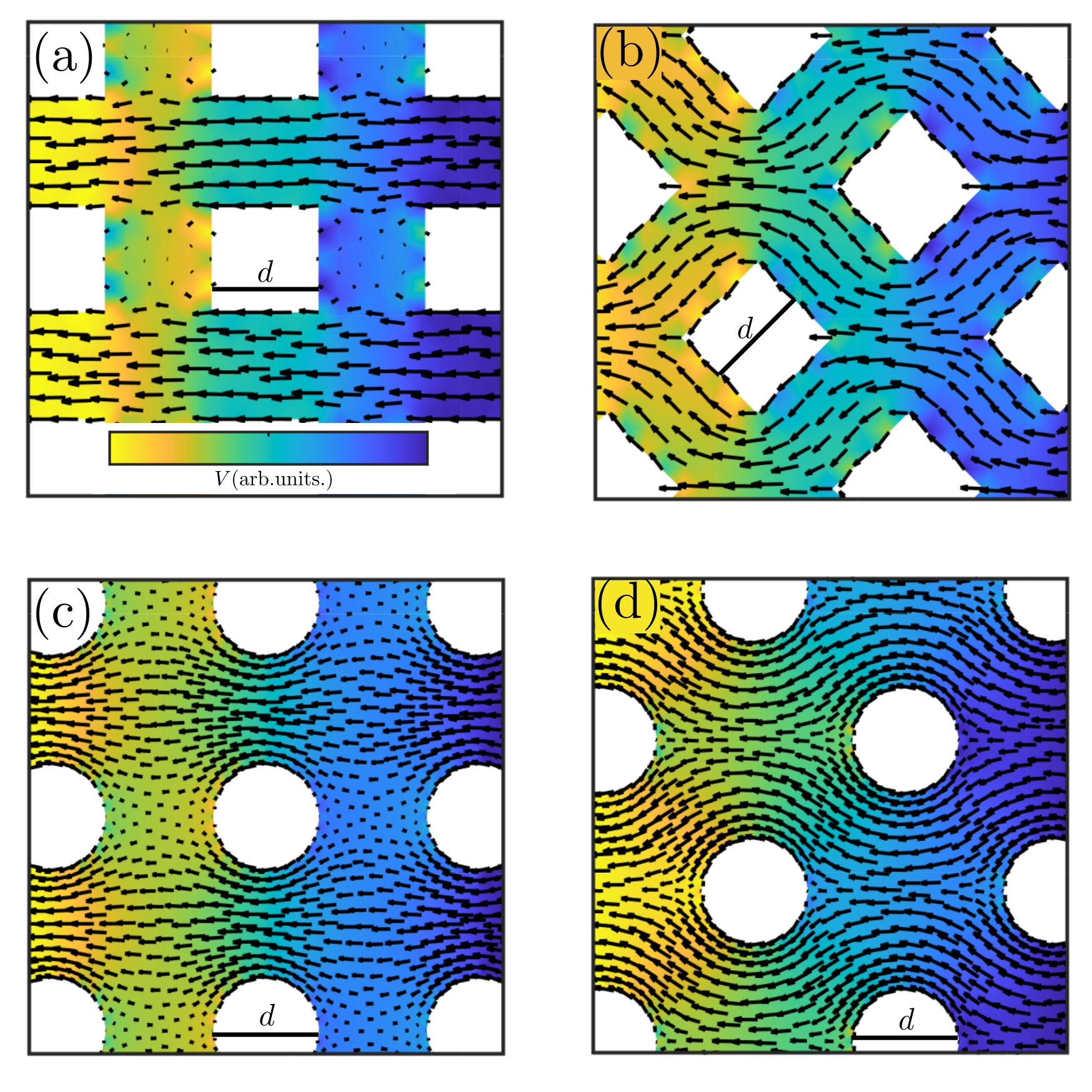}
\caption{Color maps of the electrical potential inside the sample with the streamlines for the electron fluid to visualize the bending of electron flow in antidot superlattices. Square antidots of side $d$ with a separation of $d$ yields (a) $R_1/R_2 = 2.46$ in an aligned lattice and (b) $R_1/R_2 = 2.52$ in a $45^{\circ}$ rotated lattice with respect to the current flow. Circular antidots of diameter $d$ separated by $d$ yields (c) $R_1/R_2 = 2.58$ in an aligned lattice and (d) $R_1/R_2 = 2.63$ in a $45^{\circ}$ rotated lattice with respect to the current flow.}
\label{fig:optimization}    
\end{figure}

The geometry of a device determines its hydro\-dynamic properties. Indeed, bending the electron flow enhances the hydrodynamic signatures. Figure \ref{fig:optimization} shows four antidot superlattices with different configurations and hole shapes, where the device current flows from left to right. Numerical simulations were performed with a hydrodynamic model based on the Navier-Stokes equation~\cite{meta_hydrodynamic_routes_to_viscous_electron_flow} for the characteristic lengths $l_e = 6 d$ and $l_{ee} = d$, and a perfect slip boundary condition, with a slip length $\xi \to \infty$, which corresponds to specular reflections of the electrons. In order to quantify the bending of the electron flow in every superlattice, we considered the scaling laws analyzed in Sec.~\ref{sec:results}, and evaluated the ratio $R_1 / R_2$, where $R_1~(R_2)$ is the resistance of a device with antidot size $d~(2d)$ (see Appendix~\ref{app:scalingDetail} for a detailed description). Note that $R_1 / R_2 > 2$ is a hydrodynamic signature, so we look for the largest $R_1 / R_2$ as a quantitative criterion to support a maximized bending of the electron flow. Square antidots have sharp corners, so the electronic fluid follows almost straight trajectories without bending too much, as shown in Fig.~\ref{fig:optimization}(a). As a consequence $R_1 / R_2$ is small. Moreover, given the difficulties in building and simulating samples with sharp corners, we avoid using square lattices. Superlattices with circular antidots show a higher $R_1 / R_2$, especially when they are aligned at $45^{\circ}$ with respect to the current flow [see Fig.~\ref{fig:optimization}(d)]. In fact, misalignment with the latter avoids straight trajectories, which are less prone to bending around. 

\section{Sample preparation}
\label{app:SamplePrep}

The final heterostructure, consisting of a fully encapsulated single layer graphene with a graphite bottom gate was fabricated by means of the standard mechanical exfoliation on pristine crystals of hexagonal boron nitride (hBN) and graphite, subsequently deposited onto a p-doped silicon wafer with a $300 \, \rm nm$ top layer of silicon oxide. Top hBN and bottom hBN flakes had thicknesses of $8\,$nm and $60\,$nm respectively, and the graphite back gate consisted of a $15\,$nm-thick layer. The graphite bottom gate enhances charge mobility and screens undesired spurious effects arising from charged defects present in the underlying doped silicon substrate~\cite{Disorder_in_van_der_Waals_heterostructures_of_2D_materials}.  The thicknesses of these three layers of the stack were accurately measured with a  Bruker Nano  DektakXT profilometer. For the characterization of the graphene monolayer, micro-Raman spectroscopy for the single-layer material was performed. For the stacking process of the heterostructure, a polycarbonate film was fabricated and deposited on polydimethylsiloxane. The top hBN was picked up at $50-60^\circ$ and deposited on the graphene monolayer at 190$^\circ$. Employing the same technique, the hBN bottom flake was deposited onto the graphite back gate. Subsequently, the latter stack was annealed in vacuum at 350$^\circ$ to eliminate potential residues. The hBN top and graphene were finally picked up and deposited on the hBN bottom and graphite. A thorough micro-Raman map of the final heterostructure was performed to select the cleanest and defect-free region where the electron-beam lithography was performed. 

\begin{figure}
\includegraphics[width=\columnwidth]{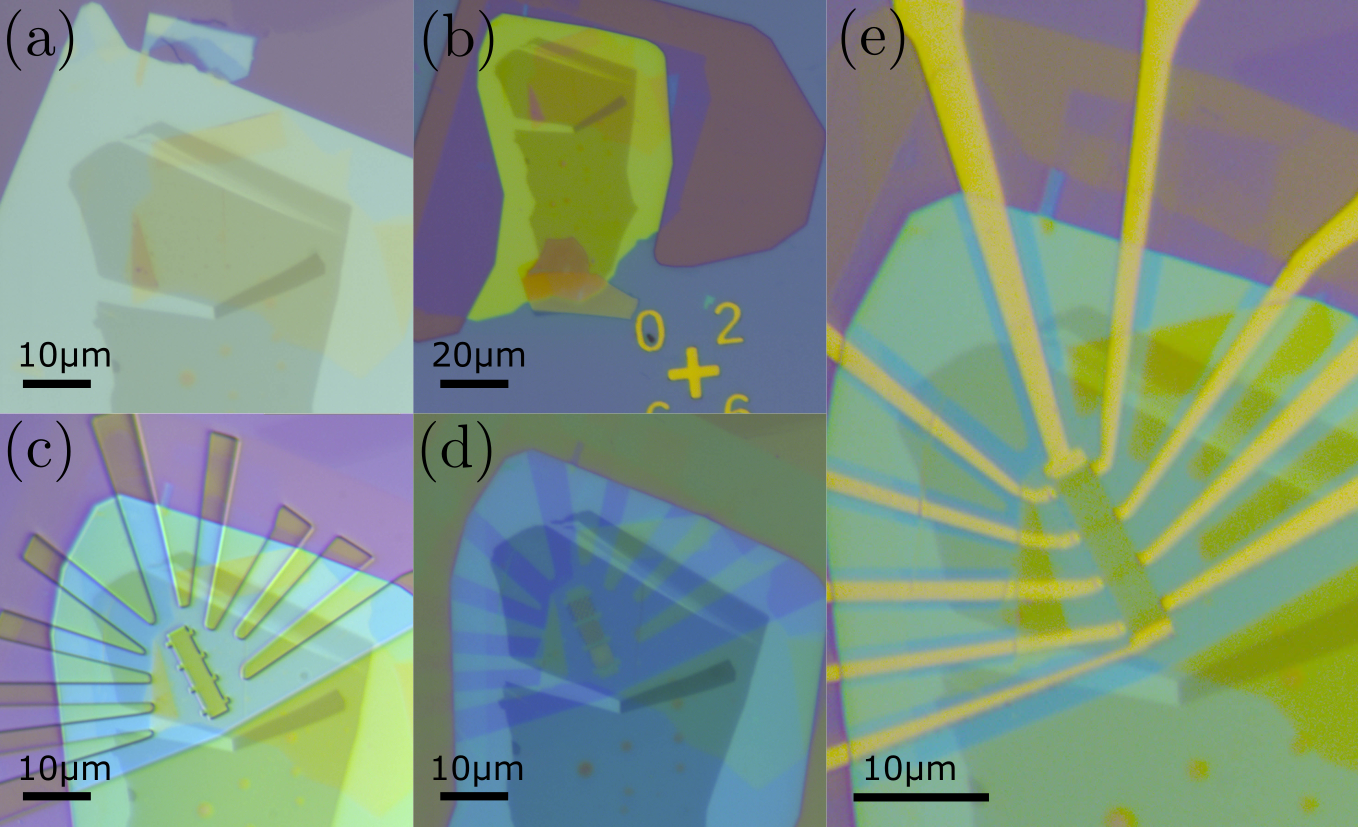}
\caption{\label{samplepreparation} Different steps in the device fabrication process and the final heterostructure (a)~the premask after etching in (b)~with the Hall bar finally defined in the heterostructure in panel (c). Panel (d) already displays the three different antidot regions with diameters of $d = 300, \, 200 $, and $100 \, \rm nm$ from the bottom region to the uppermost part of the panel. Finally, in panel (e) the final device after the evaporation of the Ohmic side contacts is shown.}
\end{figure}

Once the final heterostructure was fabricated [see Fig.~\ref{samplepreparation}(a)] a premask process was carried out by EBL-SEM to remove excess flakes around it, thus avoiding possible electric shorts between pads and facilitating further processing. A spin coating process was performed using homemade PMMA resist 5$\%$ (by weight) in chlorobenzene. The premask was attacked with a cryo-etching system [see Fig.~\ref{samplepreparation}(b)]. Then, the stack was finally shaped by means of e-beam lithography into the final 10--terminal Hall bar [see Fig.~\ref{samplepreparation}(c)]. A last step of electron beam lithography and cryo etching process was carried out to define the antidot patterns within the Hall bar [see Fig.~\ref{samplepreparation}(d)] to favor smooth edges that ensure almost specular reflection~\cite{Clerico2019,EBLvito2020}. Lastly, 10/55 nm Cr/Au contacts were deposited by e-beam evaporation [see Fig.~\ref{samplepreparation}(e)] and the device was bonded on a LCC20 chip carrier for electrical characterization, finalizing the sample fabrication.

\begin{figure}
\includegraphics[width=\columnwidth]{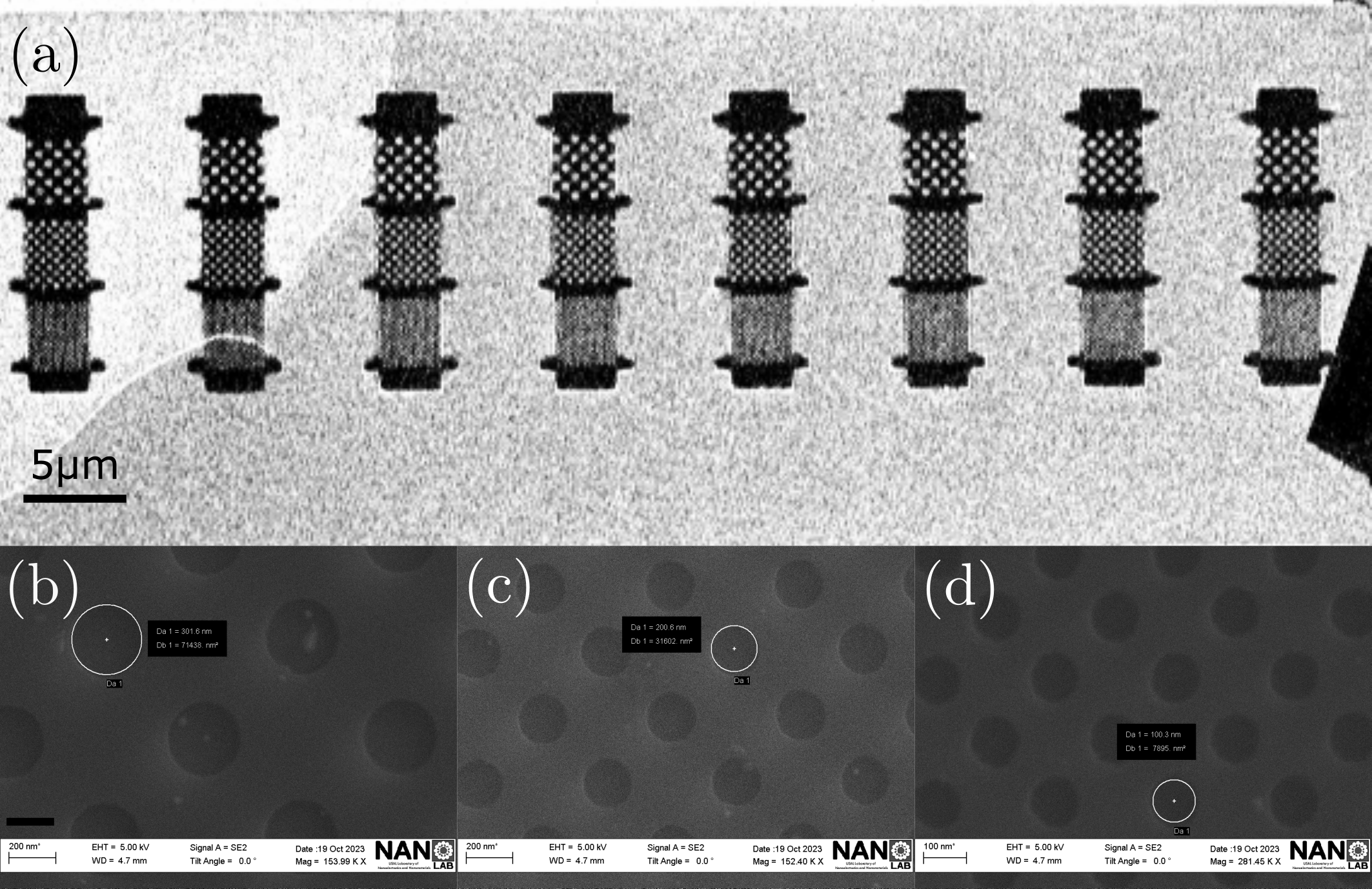}
\caption{\label{dosearray} SEM electron micrograph for the dose array optimization process. (a)~Antidot patterns obtained from doses ranging  from 400 $\mu$C/cm$^2$ (leftmost structure) to 575 $\mu$C/cm$^2$. (b)--(d)~From left to right: Antidot final structure onto an 8nm hBN flake displaying a great resolution and smooth edges for a dose of 550 $\mu$C/cm$^2$ within $d = 300, \, 200 $, and $100 \, \rm nm$ regions.}
\end{figure}

For the crucial lithographic step to define the periodic antidot lattice in the structure, a previous dose array process was carried out into a sacrificial hBN flake replicating the final design as shown in Fig.~\ref{dosearray}(a). There, doses ranging from 400 to 575 $\mu$C/cm$^2$ were used showing that 550 $\mu$C/cm$^2$ generates the optimal resolution for all three different antidot areas as shown in Figs.~\ref{dosearray}(b)--(d). 

\section{Theoretical model \label{app:theoreticalModels}}

Let us describe electrons as semiclassical particles~\cite{di_ventra_electrical_transport_in_nanoscale_systems,ballistic_and_hydrodynamic_magnetotransport_in_narrow_channels,higher_than_ballistic_conduction_of_viscous_electron_flows}, moving in a 2D device with a well-defined position ${\bm r}=(x,y)$ and wave vector ${\bm k}=k\hat{\bm u}_{\bm k}(\theta)$, where $\hat{\bm u}_{\bm k}(\theta)\equiv \left (\cos \theta , \sin \theta \right)$. Let $\bm v$ denote the velocity of the electrons. The semiclassical description ignores quantum effects but accounts for the role of the geometry in the hydrodynamic effects. Moreover, since the considered sizes $d$ are much larger than the Fermi wavelength in these devices, the impact of quantum effects is expected to be reduced~\cite{electron_trajectories_and_magnetotransport_in_nanopatterned_graphene_under_commensurability_conditions}. Then, the Boltzmann transport equation reads 
\begin{equation}
\partial_t \hat{f}  + {\bm v} \cdot \nabla_{{\bm r}}  \hat{f} - \frac{e}{\hbar} \, \big( -\nabla  \hat{V} + {\bm v} \times {\bm B} \big) \cdot \nabla_{{\bm k}}  \hat{f} =  \Gamma \left[  \hat{f} \right]\ ,
\label{BTEtimeDependent}
\end{equation}
where $\bm B$ is a perpendicular magnetic field and $\Gamma [f]$ is a collision operator
\begin{equation}
  \Gamma [f] =  \frac{f-f_e}{l_e}  + \frac{f^\mathrm{even}-f^\mathrm{even}_{ee}}{l_{ee}^\mathrm{even}} + \frac{f^\mathrm{odd}-f^\mathrm{odd}_{ee}}{l_{ee}^\mathrm{odd}} \ .
\end{equation}
This includes scattering against impurities and phonons that drive the system towards the equilibrium distribution $f_e$ with a path against impurities and phonons $l_e$ and electron-electron scattering towards an equilibrium distribution $f_{ee}$ that moves with the fluids drift velocity. We split this last term with two relaxation times for the even and odd parts of the collision operator~\cite{tomographic_dynamics_and_scale_dependent_viscosity_in_2D_electron_systems}. Thus, if $f$ is expanded in angular harmonics $f = \sum_n f_n e^{n \theta}$, $f^{\rm even}$ is the sum of the terms with $n$ even and $f^{\rm odd}$ includes the terms with $n$ odd. They have different mean free paths, namely $l_{ee}^{\rm even}$ and $l_{ee}^{\rm odd}$.  Now, let us consider an isotropic conduction band with $k_F$ the Fermi wavenumber, ${\rm v}_F$ the Fermi velocity and $m = \hbar k_F / {\rm v}_F$ the cyclotron mass. We also assume a constant carrier density $n$ such that $k_F  = \sqrt{\pi n}$, considering valley and spin degeneracy. Importantly, we assume that the relevant phenomena happen near the Fermi line, so transport can be described just in terms of the $\theta$ direction. Thus, we define
\begin{subequations}
  \begin{align}
    g({\bm r},\theta) &= \frac{\hbar}{m} \int_{0}^\infty \Big[ f({\bm r},{\bm k}) - f^e({\bm k})\Big] \, {\rm d} k \ ,  \\
    g_{ee}({\bm r},\theta) &= \frac{\hbar}{m} \int_{0}^\infty \Big[ f_{ee}({\bm r},{\bm k}) - f^e({\bm k})\Big] \, {\rm d} k \ .
    \label{eq:gee}
  \end{align}%
\end{subequations}
It is not difficult to show that $g_{ee} ({\bm r},\theta) \simeq u_x({\bm r}) \cos \theta + u_y({\bm r}) \sin \theta $, where the drift velocity is obtained as ${\bm u} ({\bm r} ) =(1/\pi) \int_0^{2\pi} g({\bm r},\theta) \hat{\bm u}(\theta)\, {\rm d} \theta$~\cite{meta_hydrodynamic_routes_to_viscous_electron_flow}.
We restrict ourselves to steady-state conditions, such that the non-equilibrium distribution function $f({\bm r},{\bm k})$ becomes independent of time. Hence, as described in Ref.~\cite{meta_hydrodynamic_routes_to_viscous_electron_flow}, the Boltzmann equation for the non-equilibrium distribution function in an electric potential $V(\bm r)$ and a perpendicular magnetic field ${\bm B}$ reduces to Eq.~\eqref{BTE}.

The geometry and edge scattering play a crucial role in viscous electron flows, so the Boltzmann equation must be supplemented with the appropriate boundary condition. Two common choices~\cite{boundary_conditions_of_viscous_electron_flow,meta_hydrodynamic_routes_to_viscous_electron_flow} are the diffusive (DF) edge that assumes $g(\theta )  = 0$ for all reflected electrons and the partially specular (PS) edge that reads
\begin{align}
g(\theta ) & = g(-\theta ) +  \mathcal{D}  \sin \theta \nonumber \\
&\times\left[ g(-\theta ) - \frac{2}{\pi}\,\sin\theta\int_{0}^{\pi} \sin^2 \theta^{\prime} g(-\theta^{\prime}) \, {\rm d} \theta^{\prime} \right]\ ,
\label{boundary}
\end{align}
where $0 < \theta < \pi$ are the reflected electrons and $-\pi < \theta < 0$ are the incident ones. For the sake of simplicity, we wrote the boundary condition for an edge parallel to $\theta = 0$. Here, $\mathcal{D} \equiv \sqrt{\pi} h^2 h'k_F^3 \lesssim 1$ is the dispersion coefficient, with $h$ the edge's bumps mean height and $h'$ its correlation length. If $\mathcal{D} \ll 1$ the edge is almost specular (see Ref.~\cite{meta_hydrodynamic_routes_to_viscous_electron_flow} for further details of the theoretical model). Indeed, we compute all the results with $\mathcal{D} = 0.01 \ll 1$, which is an almost perfect specular boundary. 

\section{Numerical methods \label{app:numericalMethods}}

We solve the Boltzmann equation numerically with a conformal Galerkin finite element method~\cite{finite_element_methods_mathematical_enigneering}. We approximate the solution as 
\begin{equation}
    g(\bm r , \theta ) = \sum_{n=1}^N \sum_{m=1}^M \phi_n (\bm r) \varphi_m(\theta) 
\end{equation}
For the spatial part, $\lbrace \phi_n \rbrace_{n=1}^N$ is the set of tent functions and the products of two tent functions defined on a triangular mesh~\cite{off_centre_steiner_points_for_delauney_refinement_on_curved_surfaces} for the antidot geometry. We impose a maximum triangle size $h < 0.1 d$ (or $h< 0.2 d$ under a magnetic field) for which $N \sim 2000 $ and convergence is ensured. For the angular part, $\lbrace \phi_m \rbrace_{n=1}^M$ is a set of periodic functions defined on the interval $[ 0 , 2\pi ) $ and we use $M = 16$ and $M = 32$. We write the weak formulation of~\eqref{BTE}, add an equation to set a uniform density of carriers and solve the resulting linear system iteratively with a least square approximation in Matlab. At the edges, we impose the boundary condition~\eqref{boundary} for reflected electrons. We solve the system on a rectangular cell of size $2 \sqrt{2} d  \times \sqrt{2} d$, and impose periodic boundary conditions. We set the potential difference between two cells across the longitudinal direction, and determine the Hall potential across the transverse direction by adding an additional equation that imposes no net current across the transverse direction. The hydrodynamic model that we used for the geometry optimization was also solved using finite elements~\cite{meta_hydrodynamic_routes_to_viscous_electron_flow}. We used a Runge-Kutta 4 method to find the electron trajectories in the streamlines, and numerical integration to find the total current. Given the carrier density $n = 0.3 \times 10^{12}\, \rm cm^{-2}$ which was determined from Hall measurements and the quantum Hall effect, the reduced units from the numerical simulation are translated into resistances. The resistances include a geometrical aspect ratio $L/W$, where $L \approx 3 \, \rm \mu m$ is the length of the region containing the antidots and $W = 3 \, \rm \mu m $ its width, so $L / W = 1$ .

\section{Sample characterization 
\label{app:mobility}}

\begin{figure}[ht!]
\includegraphics[width=\columnwidth]{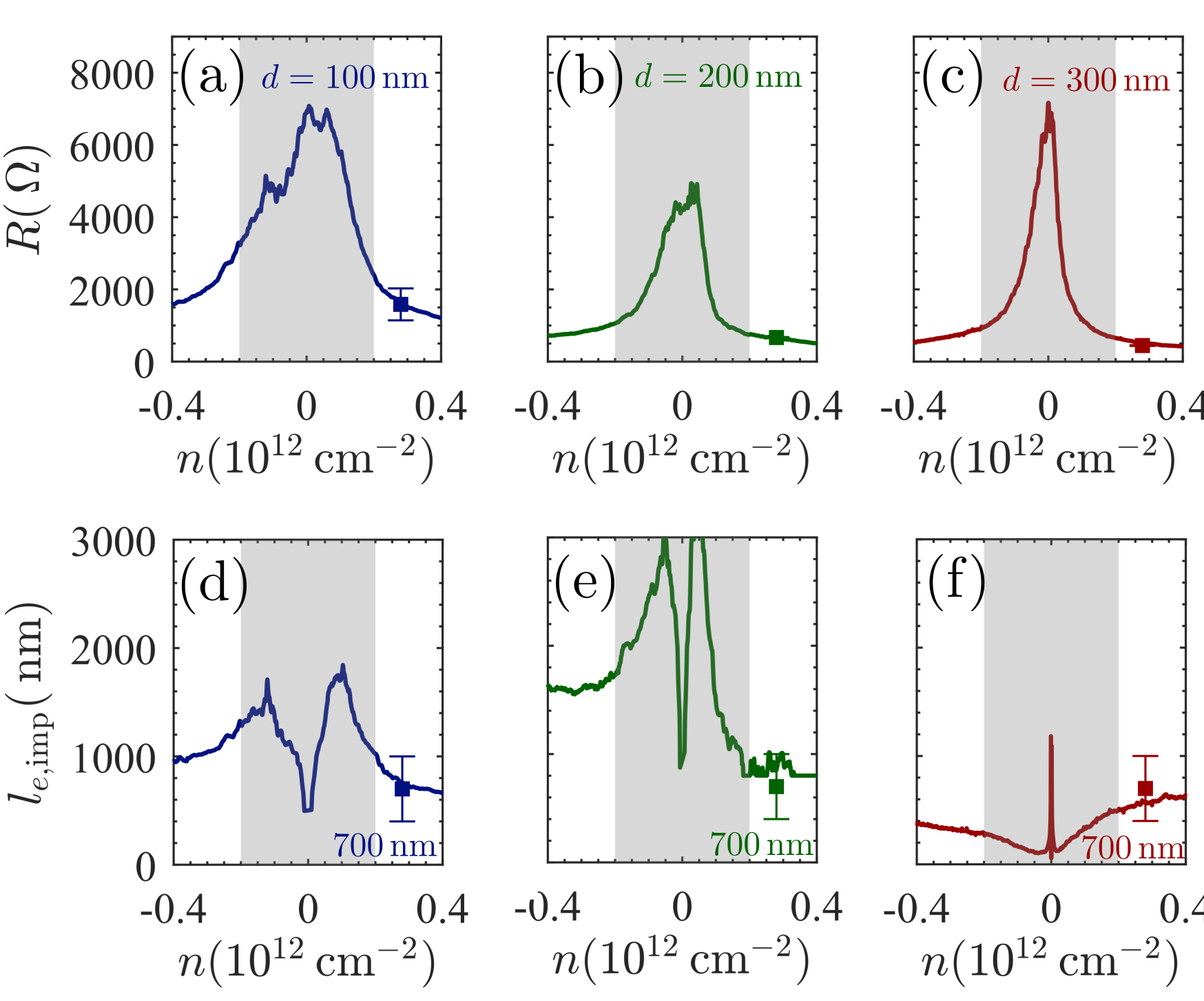}
\caption{Sample characterization. (a)-(c) Experimental Dirac peaks at $T=1.5 \, \rm K$. The nanostructuring of the device broadens the Dirac peaks (shaded area). The symbols with error bars show the simulated values of $R$ when a range of $l_{e,\rm{imp}} = 700 \pm 300 \, \mathrm {nm}$ was considered. (d)-(f) Estimation of the mean free path due to impurities $l_{e,\rm{imp}}$. The symbols with error bars indicate the range of values $l_{e,\rm{imp}} = 700 \pm 300 \, \mathrm {nm}$ considered for the numerical estimation of $R$. }
\label{fig:dirac}
\end{figure}

In this section, we present the electrical characterization of the sample. Figure~\ref{fig:dirac}(a)-(c) shows the Dirac peaks for the three regions ($d = 100, \, 200 $, and $300 \, \rm nm$) of the device at $T=1.5 \, \rm K$. The peaks broaden due to the nanostructuring of the device, which, aside from tensions near the edges, results in charge inhomogeneity (see Appendix~\ref{app:chargeInhomogenity}). Thus, we shall be careful when we discuss the results near the Dirac peaks, especially, when we compare them with simulations of the Boltzmann equation. However, this is not the case away from the charge neutrality point, where these curves show an increase of the resistance with decreasing antidot size. This means the electrical properties are strongly affected by scattering against the antidots. In particular, let us consider that the resistance at very low temperature $T\sim1.5 \, \mathrm{K}$ only depends on the antidot nanostructure and on the scattering against impurities: collisions with other electrons~\cite{Kumar17} and phonons~\cite{acoustic_phonon_scattering_limited_carrier_mobility_in_two_dimensional_extrinsic_graphene} are negligible at this temperature. Now, we have to disentangle the collisions against the antidots of size $d$ and against impurities (microscopic processes with a mean free path $l_{e,\rm{imp}}$). Notice that using a Drude-like model with Matthiessen's rule to add all collision sources $1/l_e + 1 /l_{ee} + 1 /d $ is impossible in the hydrodynamic regime, where the scattering mechanisms work together with the geometry to set the electrical properties~\cite{Gurzhi1963,Gurzhi1968}. However, we can use the Boltzmann equation to disentangle the various collision mechanisms by solving it to find the $l_{e,\rm{imp}}$ that results in the observed resistance. Figures~\ref{fig:dirac}(d)--(f) shows the estimated $l_{e,\rm{imp}}$, that gives rise to the experimental resistances of Figs.~\ref{fig:dirac}(a)--(c) by way of the Boltzmann transport simulations. We also find that the resistance is robust regardless of the particular value of $l_{e,\rm{imp}}$, and any value within the range $l_{e,\rm{imp}} = 700 \pm 300 \, \mathrm{nm}$ could explain the experimental results in Figs.~\ref{fig:dirac}(a)--(c). This is the dependence that we expect when the electrical properties are not strongly affected by impurity scattering, but rather by the antidot geometry. In particular, given the moderate influence of $l_{e,\rm{imp}}$, there is no need to consider different $l_{e,\rm{imp}}$ to account for the experimental results in the three antidots regions. In view of these results, we assume $l_{e,\rm{imp}} = 700 \, \rm nm$ at  $n = 0.3 \times 10^{12} \, \rm cm^{-2}$, similar to the one achieved in other graphene nanostructures~\cite{ballistic_transport_in_graphene_antidot_lattices}. Last, at higher temperatures $T$, we add phonon scattering $1/l_e = 1/l_{e,\rm{imp}} + 1/l_{e,\rm{ph}}$, with a Bloch-Gr\"uneisen temperature of $54 \, \rm K$ to account for the low-temperature non-linear dependence~\cite{acoustic_phonon_scattering_limited_carrier_mobility_in_two_dimensional_extrinsic_graphene}. This assumes that phonons do not affect the electrical properties below $\approx 10 \, \rm K$ and a high-temperature linear dependence of $l_{e,\mathrm{ph}} \approx 170 \, {\rm \mu m \, K} /T$~\cite{phonon_mediated_room_temperature_quantum_hall_transport_in_graphene}.

\section{Charge inhomogeneity \label{app:chargeInhomogenity}}

\begin{figure}[t!]
\includegraphics[width=\columnwidth]{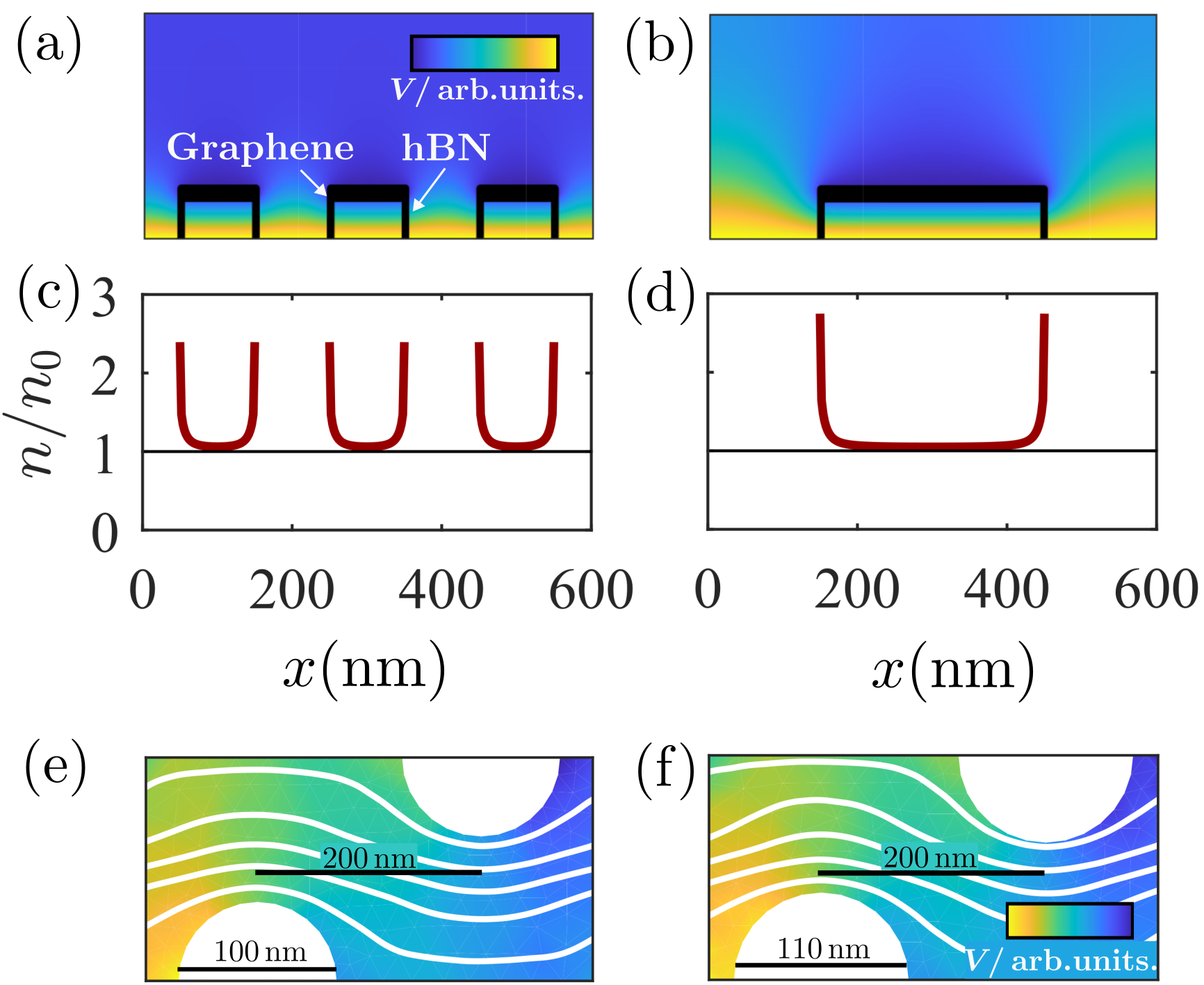}
\caption{Charge inhomogeneity in antidot superlattices. (a)-(b) Transverse section of a graphene conductor, encapsulated between two layers of hBN, on top of a back gate at a given potential, for $d = 100 $ and $300\, \rm nm$ hole sizes. (c)-(d) Induced density of carriers in the flake, normalized to the $n_0$ density corresponding to an ideal capacitor. (e)-(f) Color maps of the electrical potential inside the sample with the streamlines for the electron fluid simulated with the Boltzmann transport equation. Left (right) panels show results  for antidots of diameter $d$ ($1.1 d$).}
\label{fig:chargePlate}
\end{figure}

We engineer the geometry of the device in order to bend the electron flow. As a side effect, the back gate induces an inhomogeneous density of carriers $n$. Let us quantify the underlying electrostatic effect for a graphene flake encapsulated between two layers of hBN, a dielectric medium with $\varepsilon \sim 3.5$, with the top and bottom thicknesses of $8\, \rm nm$ and $60 \, \rm nm$ respectively. For this purpose, we consider a simplified two dimensional setup by assuming that the antidots extend across the direction perpendicular to the figure, where we can solve the Poisson equation with a finite differences numerical method. Figure~\ref{fig:chargePlate} shows the solution: panels~(a) and~(b) show the electrostatic potential and panels~(c) and~(d) account for the induced density of carriers. Indeed, the carrier density is not homogeneous, and the accumulation next to the edges would be more relevant for superlattices of shorter $d$. 

Charge inhomogeneity broadens the Dirac peak of geometrically engineered samples, shortens the effective mean free paths $l_e$, and leads to small discrepancies when the same value of $l_e$ is used for all antidot regions. We quantify the effect of charge inhomogeneity by studying the most dramatic scenario where the region adjacent to the antidots does not contribute to electrical conduction, for example, when it is depleted of free carriers. Therefore, we can simply simulate a system with the same center-to-center distance, but with a bigger antidot diameter. The results for archetypal experimental parameters at low temperatures and $d/l_e = 0.15$ are shown in panels~(e) and~(f). We find that even a major change of $10 \%$ in the antidot diameter only results in a change of $17\%$ in the resistance. Although it is quite remarkable, it is not enough to explain the most relevant results of our work.

Last, we notice that, due to the doping of the graphene flake with holes, the applied voltage needed to achieve a $-n$ density of holes is smaller, in modulus, than to achieve the same density of electrons $n$. Thus, the inhomogeneity is less noticeable when working with holes. This may explain the better agreement with the simulations when working with holes [see Fig.~\ref{fig:superballistic}(b) and~(c)]. Last, if the scaling laws were caused by the inhomogeneity, they should be dramatically different for electrons and holes. Consequently, the fact that the scaling $R_1 / R_2 > 2$ prevails both for types of carriers [see Fig.~\ref{fig:scaling}(a)] shows that it is not due to charge inhomogeneity. 

\section{Charge neutrality \label{app:chargeNeutrality}}

\begin{figure}[ht!]
\includegraphics[width=\columnwidth]{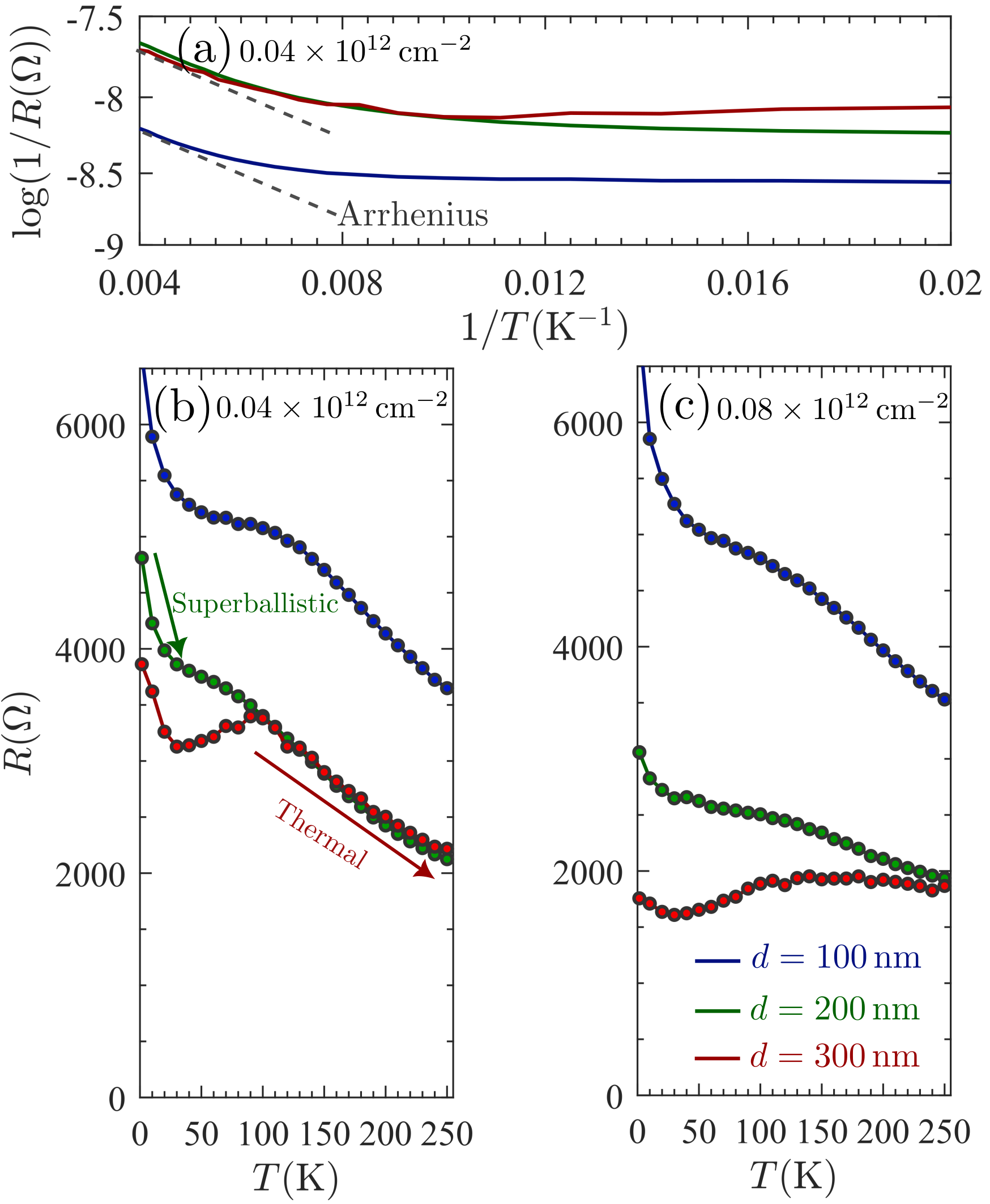}
\caption{Resistance measurements near the charge neutrality point. (a)~Arrhenius's law characterizes thermally activated conduction. (b)-(c) Resistance as a function of the temperature for densities of carriers.  }
\label{fig:neutrality}
\end{figure}

Superballistic conduction is often studied far away from the charge neutrality point~\cite{Kumar17,geometric_control_of_universal_hydodynamic_flow_in_a_two_dimensional_electron_fluid,long_distance_electron_electron_scattering_detected_with_point_contacts}. Conversely, in this section, we will focus on superballistic conduction near the charge neutrality point. Figure~\ref{fig:neutrality}(a) shows an Arrhenius plot of the resistance as a function of temperature near the charge neutrality point with two well-differentiated physical regimes. The high-temperature region is dominated by non-hydrodynamic thermal effects and so, the experimental data does fit the Arrhenius law. For low temperatures $T \lesssim 100 \, \rm K$~\cite{phonon_mediated_room_temperature_quantum_hall_transport_in_graphene}, this is not the case as a result of the collective electron flow. Indeed, Figs.~\ref{fig:neutrality}(b) and (c) (both for a different carrier density) show two distinguishable steps in the experimental resistance: superballistic conduction occurs at $T \lesssim 100 \, \rm K$ and other thermal phenomena $T \gtrsim 100 \, \rm K$. Notice that thermal excitations present at higher temperatures boost the transition from the inhomogeneity regime to the regime of Dirac plasma~\cite{giant_magnetoresistance_of_dirac_plasma_in_high_mobility_graphene}. Other thermal processes given in intrinsic semiconductors or insulators result in a similar behavior ($\mathrm{d} R / \mathrm{d} T$<0).
Most importantly the decrease of resistance due to the superballistic conduction or the Dirac plasma effect occurs at distinguishable temperatures and has a different functional dependency~\cite{giant_magnetoresistance_of_dirac_plasma_in_high_mobility_graphene,Kumar17}.
 The fact that the descent at low temperature ($T \lesssim 50 \, \rm K$) is highly dependent on $d$ further shows that it is a geometrical effect, as hydrodynamic theory predicts, and it can be mainly attributed to superballistic conduction.

\section{Boundary scattering and universality \label{app:boundaryScattering}}

One of the main questions regarding electron hydrodynamics is that of the edge scattering, which determines electrical properties to some extent~\cite {boundary_conditions_of_viscous_electron_flow}. The cryo-etching technique gives control of the graphene edge, with bumps of mean height $h$ and correlation lengths $h'$ in the order of the nm~\cite{Clerico2019,EBLvito2020}. Since $k_F^{-1} \gg 1 \, \rm nm$ unless we are next to the Dirac peak, we assume $\mathcal{D} = \sqrt{\pi} h^2 h' k_F^3 \ll 1$ for the dispersion coefficient in the simulations. Indeed, we compute all the results with $\mathcal{D} = 0.01 \ll 1$, which is practically perfectly specular. In this sense, this is the same as the boundary condition for electrostatically defined edges in GaAs heterostructures~\cite{geometric_control_of_universal_hydodynamic_flow_in_a_two_dimensional_electron_fluid}, providing a mechanism for the control of edge scattering in graphene, whose gaplessness does not allow electrostatically defined edges. This provides results in agreement with the experiment. 

\begin{figure}[ht!]
\includegraphics[width=\columnwidth]{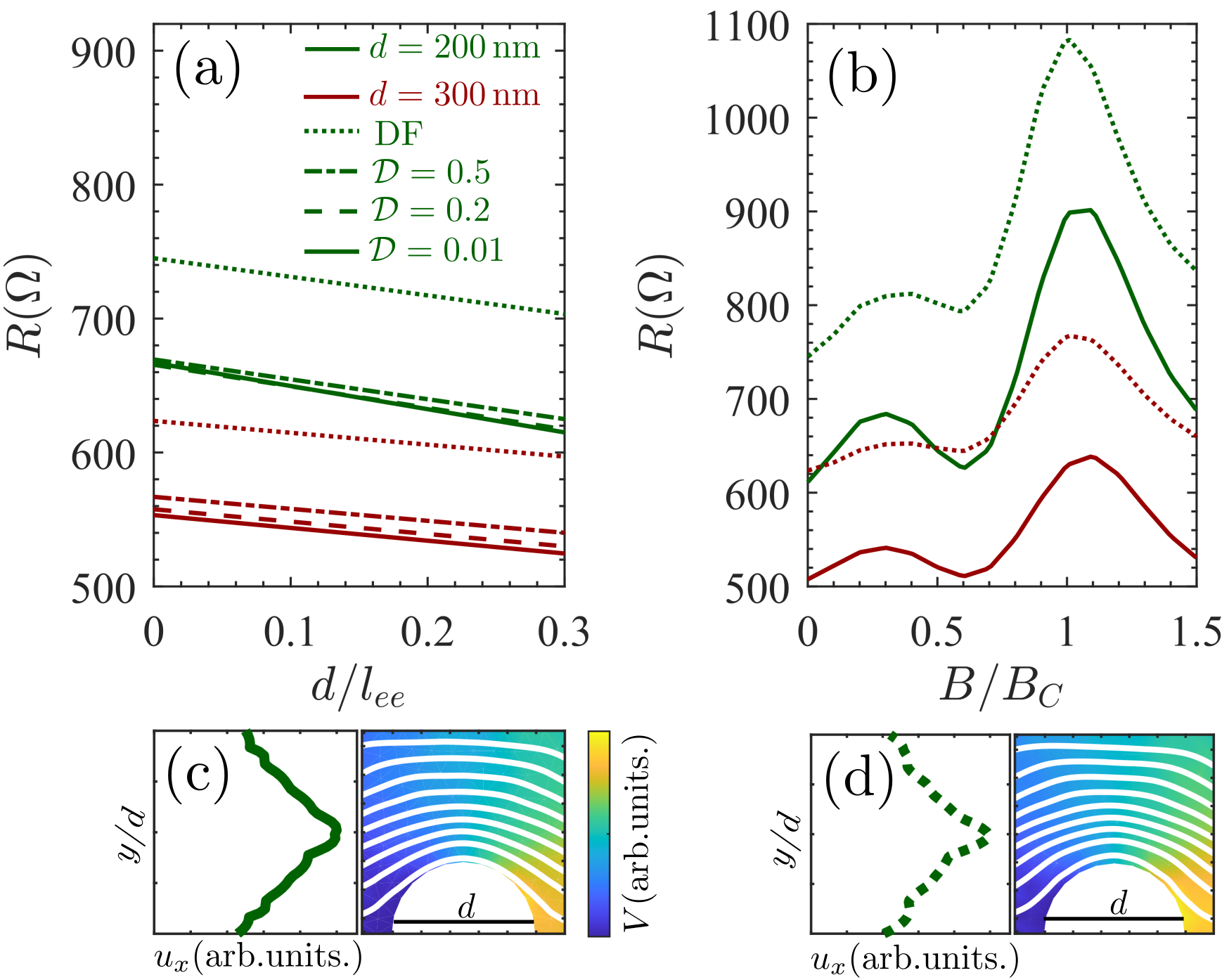}
\caption{Robustness of the results regardless of boundary scattering. (a)~Simulations of the Boltzmann transport equation for the resistance as a function of the electron-electron collision rate $d/l_{ee}$, being $l_e = 700 \rm \, nm$. We show the result for a diffusive (DF) edge and several values of $\mathcal{D}$ in a partially specular edge. (b)~Simulations of the magnetoresistance for some cases are shown in (a). (c)--(d)~Streamlines and velocity profiles for a specular boundary with $\mathcal{D} = 0.01$ and a diffusive one, for $l_e = 3 d $, $l_{ee} \gg d$.}
\label{fig:boundary}
\end{figure}

In this section, we investigate the role of edge scattering by performing simulations of the Boltzmann transport equation. Figures~\ref{fig:boundary}(a)--(b) shows the resistance as a function of the electron-electron scattering rate and the magnetic field, for several boundary conditions. The result is almost constant for several values of $\mathcal{D}$ in a partial slip boundary condition. A specular boundary condition is shown in Fig.~\ref{fig:boundary}(c), where the electron streamlines perfectly follow the shape of the holes. The electrical properties are not very different for the diffusive edge, whose velocity profile is shown in Fig.~\ref{fig:boundary}, and whose streamlines are slightly separated from the antidots. Consequently, we prove that the results are robust regardless of the edge scattering mechanism. Both the edge control of the cryo-etching technique and the fact that the electron bending is mainly controlled by the antidot geometry and not by edge scattering, ensure an almost universal viscous electron flow in our graphene antidot superlattices. 

\section{Reproducibility \label{app:reproducibility}}

\begin{figure*}[ht!]
\includegraphics[width=18cm]{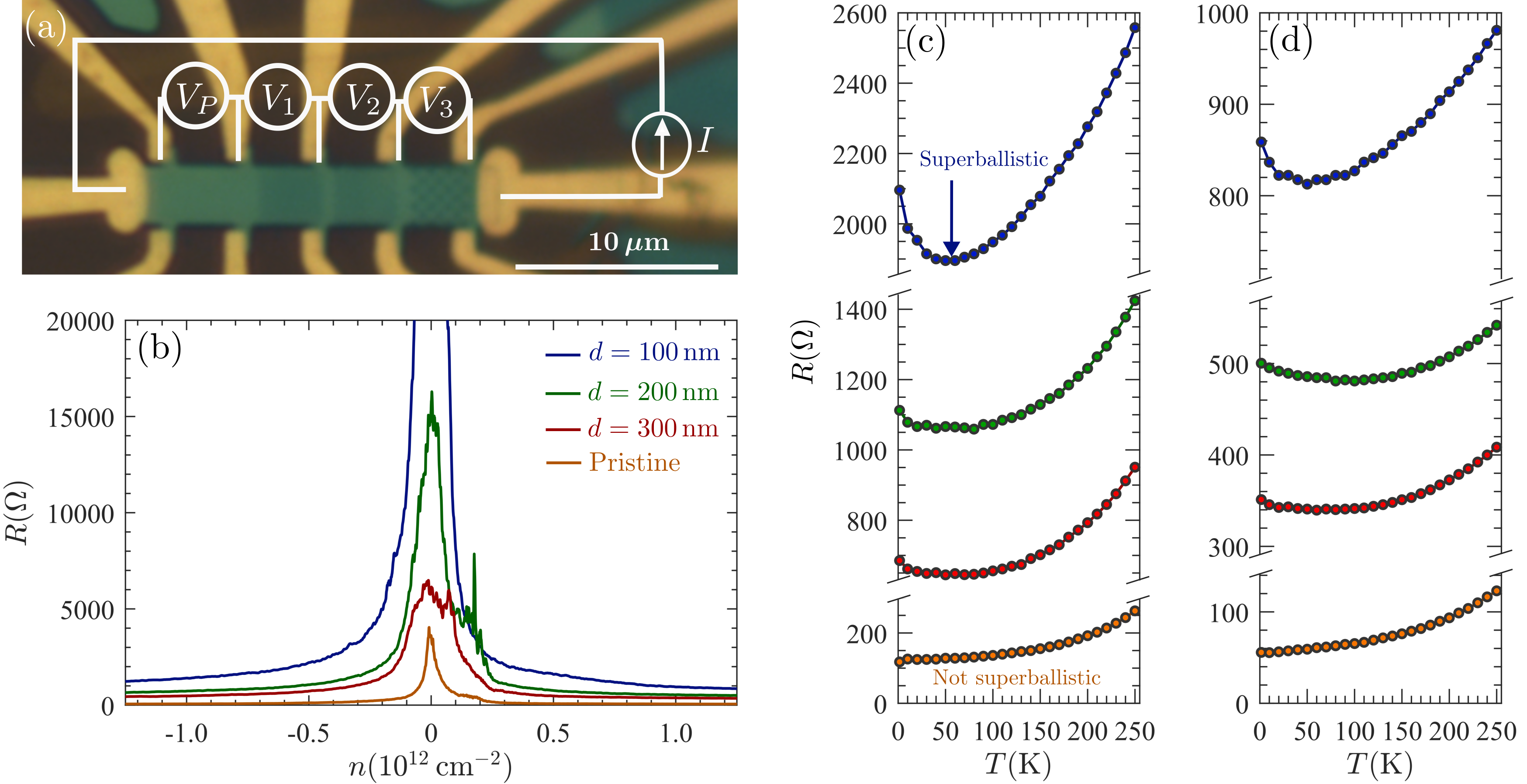}
\caption{Superballistic conduction only arises in regions with antidots. (a)~Optical image of the second superlattice, with a pristine region and antidots of diameters $d = 100, \, 200 $, and $300 \, \rm nm$. (b) Dirac peaks in all the samples. (c)--(d) Resistance as a function of the temperature for the densities of carriers $n = 0.3 \times 10^{12} \rm \, cm^{-2}$ and  $n = 1.2 \times 10^{12} \rm \, cm^{-2}$, respectively.    }
\label{fig:reproducibility}
\end{figure*}

In order to further support our findings, we fabricated a second device. The new Hall bar includes the same three regions with antidots, as well as a pristine Hall-bar-like region where no antidots were defined. The experimental procedure is described in Appendix~\ref{app:SamplePrep}, but now a standard 300-nm-thick SiO$_2$ substrate was used to enable higher gate voltages without dielectric breaking. While the top hBN flake was kept comparable to the one used for the graphite back-gate sample ($\sim10\,$nm), the bottom one was significantly thinner ($\sim 30\,$nm) since there was no risk of electrical short from the graphene towards an underlying metallic layer (graphite). Figure~\ref{fig:reproducibility}(a) shows the Dirac peaks of the new sample, with the resistance as a function of carrier density. Figures~\ref{fig:reproducibility}(b) and (c) shows the superballistic conduction in the regions with antidots, being qualitatively similar to the ones studied in Fig.~\ref{fig:superballistic} for different densities of carriers. Most importantly, no superballistic effect arises in the pristine region. In conclusion, the values of ${\rm d } R / {\rm d} T < 0 $  in the $d = 100, \, 200 $, and $300 \, \rm nm$ regions, further justify the finding of superballistic conduction in antidot superlattices. Moreover, the absence of the effect in the pristine region with no antidots suggests that the geometrical engineering of the device is responsible for the superballistic conduction. 

\section{Tomographic and hydrodynamic approaches \label{app:tomographic}}

\begin{figure}[ht!]
\includegraphics[width=\columnwidth]{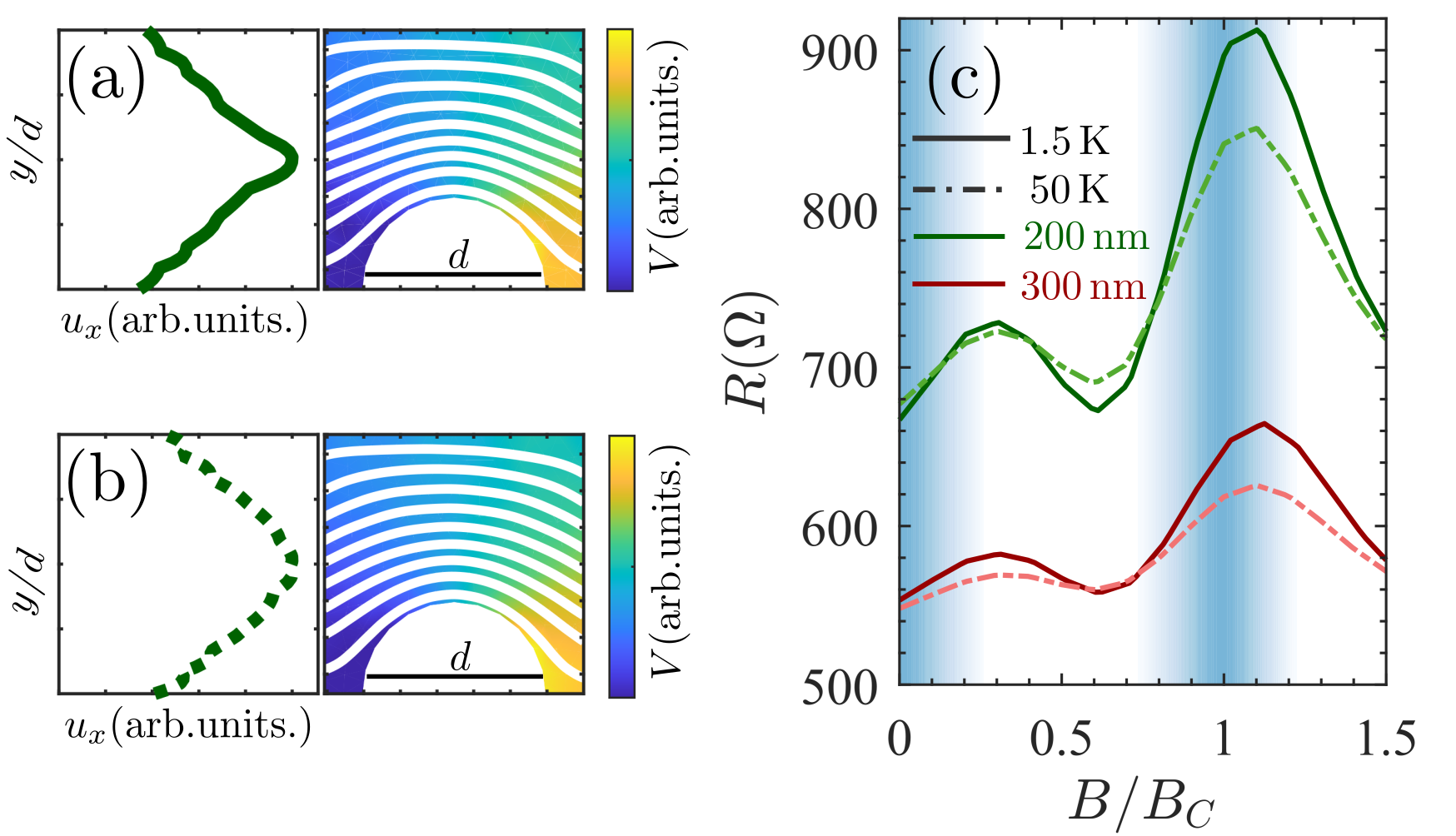}
\caption{Tomographic and hydrodynamic transport. (a)~Simulations of the Boltzmann equation for electron streamlines and current profiles, being $d / l_e = 0.25$, $d/l_{ee}^{\rm even} = 0.4$ and $d/l_{ee}^{\rm odd} = 0$ in the tomographic approach followed in this paper. (b)~Same with $d/l_{ee}^{\rm even} = d/l_{ee}^{\rm odd}  = 0.4$ and $d/l_{ee}^{\rm odd} = 0$ in a fully hydrodynamic description. (c)~Magnetoresistance simulations in the fully hydrodynamic approach.}
\label{fig:tomographic}
\end{figure}

We work under the tomographic approach~\cite{tomographic_dynamics_and_scale_dependent_viscosity_in_2D_electron_systems,linear_in_temperature_conductance_in_electron_hydrodynamics,collective_modes_in_interacting_two_dimensional_tomographic_fermi_liquids,anomalously_long_lifetimes_in_two_dimensional_fermi_liquids,nonequilibrium_relexation_and_odd_even_effect_in_finite_temperature_electron_gases}, where $l_{ee}^\mathrm{even} = l_{ee}$ and $l_{ee}^\mathrm{odd}\gg l_{ee}$. Let us compare it with the conventional hydrodynamic description $l_{ee}^\mathrm{even} = l_{ee}^\mathrm{odd} = l_{ee}$. This comparison is possible by going beyond the approximated anti-Matthiessen rule~\cite{higher_than_ballistic_conduction_of_viscous_electron_flows,Kumar17} and solving the Boltzmann equation.
Notice that, despite the tomographic approach for electron-electron collisions, collisions against impurities and phonons in $l_e$ contribute equally to the even and odd-parity modes~\cite{viscosity_of_two_dimensional_electrons}. Consequently, some features of a fully tomographic regime may blur due to impurities. Still, we can investigate the differences between the hydrodynamic and tomographic approaches. Figure~\ref{fig:tomographic}(a) and (b) show the current distribution under both approaches. Also, Fig.~\ref{fig:tomographic}(c) shows the magnetic response under a purely hydrodynamic description, to be compared with Fig.~\ref{fig:intermittent}(d) for the tomographic description. As the magnetic field increases, it rotates the velocity of the electrons and makes the distinction between even and odd parity modes less noticeable. However, the behavior is different in the absence of a magnetic field. Therefore, simulations show a difference between the electrical response depending on the microscopic scattering mechanisms.   

\section{Scaling laws \label{app:scalingDetail}}

Hydrodynamic flow features properties that strongly depend on the scale. Let us analyze some relevant scaling laws affecting the device resistance in the hydrodynamic regime, $R\propto 1/d^\alpha$. In the latter, the Boltzmann equation can be reduced to a modified Navier-Stokes equation together with the continuity equation as follows~\cite{meta_hydrodynamic_routes_to_viscous_electron_flow}
\begin{align}
\nabla \cdot {\bm u} = & 0 \\
    - \nu \nabla^2 {\bm u} + \frac{{\rm v}_F}{l_e} {\bm u} = & \frac{e}{m} \nabla V
    \label{NSEsimplifiedForScaling}
\end{align}
where $\bm u$ is the fluid drift velocity and $\nu = {\rm v}_F \cdot (l_e^{-1}+l_{ee}^{-1} )^{-1} / 4$ is the viscosity. For simplicity, we write the equation in the absence of a magnetic field. Equation~\eqref{NSEsimplifiedForScaling} includes a dissipative term that accounts for collisions against impurities and phonons. These equations are solved for a particular boundary condition, that in our experimental setup describes a perfectly specular edge. This imposes the no-trespassing $u_{\perp} = 0$ and the $\partial_\perp u_{\parallel} = 0$ conditions of perfect slip, corresponding to specular edges, where $u_\parallel $ and $u_\perp$ are the components of the velocity parallel and perpendicular to the edge. 

Let us first study the physical situation such that the viscous term dominates
\begin{equation}
    -\nu \nabla^2 {\bm u_V}  = \frac{e}{m} \nabla V_V \ .
    \label{NSEviscous}
\end{equation}
In this case the solution for any geometrical size $d$ reads as ${\bm u_V} ({\bm r}) = {\bm {\tilde u_V}} ({\bm r}/d) $ and ${V_V} ({\bm r}) = d^{-1} \tilde{V}_V ({\bm r}/d)$, where $\tilde{\bm u}_V$ and $\tilde{V}_V$ are functions that do not depend on $d$. Notice that these functions fulfill the specular boundary condition. Furthermore, although this reasoning would not be valid for partial slip boundary conditions, it also applies to the no-slip $u_\parallel = 0$ condition commonly used for the derivation of the Poiseuille law in conventional fluids. As a consequence of the scale dependence, $V_V({\bm r}) \propto 1/d$ is the voltage drop in a region of length $d$, and the resistance for devices of the same geometry and different sizes scales as $R_V \propto 1/ d^2$, i.e. $\alpha = 2$. 

On the contrary, if the diffusive term dominates, Eq.~\eqref{NSEsimplifiedForScaling} reduces to
\begin{equation}
\frac{{\rm v}_F}{l_e} {\bm u_D} = \frac{e}{m} \nabla V_D \ ,
\label{NSEdiff}
\end{equation}
which is solved by the family of solutions ${\bm u_D} ({\bm r}) = {\bm {\tilde u}_D} ({\bm r}/d) $ and ${V_D} ({\bm r}) = d~\tilde{V}_D ({\bm r}/d)$. Regardless of the boundary condition, this results in a constant resistance $R_D$, i.e. $\alpha = 0$.

In the general scenario considered in Eq.~\eqref{NSEsimplifiedForScaling}, there is no trivial expression for $\alpha$. However, we propose to make the following ansatz: ${\bm u_V} ({\bm r} )={\bm u_D} ({\bm r} )={\bm u} ({\bm r} )$. Namely, there is a single velocity field ${\bm u} ({\bm r} )$ that solves both the equation with the viscous term and the dissipative term. However, the associated potentials and resistances may not be the same, so we define the bending coefficient $\beta= R_{V}/R_{D}$. The latter only depends on the geometry such that a higher $\beta$ corresponds to a more irregular electron flow. In order to estimate $\beta$ we consider the total resistance in Eq.~\eqref{NSEsimplifiedForScaling} as proportional to the sum of two terms (associated with the limiting cases considered in Eqs.~\ref{NSEviscous} and~\eqref{NSEdiff} that are independently simulated
\begin{equation}
    R \propto \frac{\beta\nu}{{\rm v}_F d^2} + \frac{1}{l_e} \ .
\end{equation}
yielding Eq.~\eqref{alphaMainArticle}.
A similar expression was derived for the averaged resistance for 2D GaAs samples, where large-radius oval defects may arise in the process of fabrication~\cite{negative_magnetoresistance_in_viscous_flow_of_two_dimensional_electrons}.
Here, we estimate $\beta\simeq 4.4$ for the antidot geometry.
Last, notice that this expression is consistent with an exponent $0 < \alpha < 2$.



\bibliography{\jobname.bib} 

\end{document}